\documentclass[useAMS, usenatbib]{mn2e}
\usepackage{times}
\usepackage{amsmath}
\usepackage{xspace}
\usepackage{psfig}
\usepackage{defs}

\def\eq{equation}
\def\fig{Fig.}

\def\tab{table}

\def\cf{{\it cf.}}
\def\ie{{\it i.e.}}
\def\eg{{\it e.g.}}
\def\ltsima{$\; \buildrel < \over \sim \;$}
\def\simlt{\lower.5ex\hbox{\ltsima}}
\def\gtsima{$\; \buildrel > \over \sim \;$}
\def\simgt{\lower.5ex\hbox{\gtsima}}
\def\fesc{{$\langle f_{\rm esc}\rangle$}\xspace}
\def\h2{H$_2$\xspace}
\def\H2{H$_2$\xspace}
\def\m{$^{-1}$\xspace}

\def\mmm{$^{-3}$\xspace}
\def\pp{$^2$\xspace}

\def\ion#1#2{\text{#1\,\sc #2}}
\def\HI{{\ion{H}{i} }}
\def\HII{{\ion{H}{ii} }}
\def\GI{{\ion{He}{i} }}
\def\GII{{\ion{He}{ii} }}

\def\popII{population~II\xspace}
\def\pop3{population~III\xspace}

\def\p3{``small-halo''\xspace}
\def\pp3{``Small-halo''\xspace}

\def\Mpc{h$^{-1}$ Mpc\xspace}

\def\Ms{$h^{-1}$~M$_\odot$\xspace}

\def\lya{Lyman-$\alpha$\xspace}

\def\taue{$\tau_{\rm e}$\xspace}
\def\sig8{$\sigma_8$\xspace}
\def\zrei{$z_{\rm rei}$\xspace}
\def\euv{$\epsilon_{\rm UV}$\xspace}
\def\xe{$\langle x_{\rm e} \rangle$\xspace}
\def\bi{\begin{itemize}}
\def\ei{\end{itemize}}

\title[Reionisation, chemical enrichment and
  seed black holes from the first stars]{Reionisation, chemical enrichment and
  seed black holes~\\ from the first stars: is Population~III important?}

\author[M. Ricotti and J.P. Ostriker]{M. Ricotti and J.P. Ostriker\\ 
Institute of Astronomy, Madingley Road, Cambridge CB3 0HA\\
ricotti@ast.cam.ac.uk, jpo@ast.cam.ac.uk}
\date{Accepted ---. Received ---; in original form 10 December 2002}
\pagerange{\pageref{firstpage}--\pageref{lastpage}}
\pubyear{2002}
\begin{document}
\maketitle
\label{firstpage}

\begin{abstract}
  We investigate the effects of a top-heavy stellar initial mass
  function on the reionisation history of the intergalactic medium
  (IGM). We use cosmological simulations that include
  self-consistently the feedback from ionising radiation, H$_2$
  dissociating radiation and supernova (SN) explosions.  We run a set
  of simulations to check the numerical convergence and the effect of
  mechanical energy input from SNe. In agreement with other studies we
  find that it is difficult to reionise the IGM at \zrei$>10$ with
  stellar sources even after making extreme assumptions. If star
  formation in $10^9$ M$_\odot$ galaxies is not suppressed by SN
  explosions, the optical depth to Thomson scattering is \taue$\simlt
  0.13$. If we allow for the normal energy input from SNe or if
  pair-instability SNe are dominant, we find \taue$\simlt 0.09$.
  Assuming normal yields for the first stars (\pop3), the mean
  metallicity of the IGM is already $Z/Z_\odot=2 \times 10^{-3}$
  ($10^{-3}<Z/Z_\odot <1$ in overdense regions) when the IGM mean
  ionisation fraction is less than 10 \%.  For these reasons \pop3
  stars cannot contribute significantly to reionisation unless the
  mechanical energy input from SNe is greatly reduced and either the
  metal yield or the mixing efficiency is reduced by a factor of
  $10^3$. Both problems have a solution if \pop3 stars collapse to
  black holes. This can happen if, having masses $M_*< 130$ M$_\odot$,
  they are characterised by heavy element fall-back or if, having
  masses $M_*>260$ M$_\odot$, they collapse directly onto black holes
  without exploding as SNe.  If metal-poor stars are initially
  important and collapse to black holes is the typical outcome, then
  the secondary emission of ionising radiation from accretion on SN
  induced seed black holes, might be more important than the primary
  emission.
  
  We also develop a semianalytic code to study how \taue is sensitive
  to cosmological parameters finding essentially the same results.
  Neglecting feedback effects, we find simple relationships for \taue
  as a function of the power spectrum spectral index and the emission
  efficiency of ionising radiation for cold dark matter and warm dark
  matter cosmologies. Surprisingly, we estimate that a warm dark
  matter scenario (with particle mass of 1.25 keV) reduces \taue by
  only approximately 10\%.
\end{abstract}

\section{Introduction}\label{sec:int}
It is widely believed that the earliest generations of very low
metallicity stars would have been far more effective than normal stars
in reionising the universe, so much so that they could account for the
large optical depth to Thompson scattering, \taue$=0.17 \pm 0.04$,
measured by the WMAP satellite \citep{Bennet:03, Kogut:03}. The main
reason for this expectation is that the stellar initial mass function
(IMF) is thought to be tipped towards high masses for $Z/Z_\odot \ll
10^{-4}$.  But there are two byproducts which are likely to follow
from this scenario. The same stars, which are efficient UV producers,
will die as supernovae (SNe), the explosions heating the surroundings
and inhibiting further star formation (``negative feedback''), and the
metals ejected in these explosions will contaminate the high density
regions, rapidly bringing the metallicity up to a level where such
stars cannot form. We argue that, while it is possible to evade these
strictures, it is very difficult to do so unless this early generation
ends its life primarily {\it via} implosion to black holes rather than
explosion as supernovae, and that in the former case the secondary
effects of ionising radiation from accretion onto the formed seed
black holes may dominate over the primary UV from the first generation
stars.

In this discussion we are focusing on the period $15<z<25$ when the
haloes within which star formation would occur are small and gas
cooling is initially via \H2 \citep{CouchmanR:86, OstrikerG:96}. The
later phase, responsible for the more complete reionisation of the
universe as attested to by the Sloan Digital Sky Survey (SDSS) quasars
at $6<z<7$ \citep{Becker:01, Fan:03}, is fairly well understood and
can easily be accounted for by a relatively normal \popII. For a
Salpeter mass function of \popII stars in the mass range $1 {\rm
  M}_\odot \le {\rm M}_* \le 100 {\rm M}_\odot$, and an escape
fraction of UV photons from the forming galaxies to the IGM of
\fesc$=20-50$\%, \cite{Gnedin:00, CiardiSW:03, Ricotti:03, ChiuFO:03}
among others, find little difficulty in matching the QSO-determined
ionisation history using standard stellar interiors theory and
cosmological models.  The two epochs of ionisation might have been
distinct, as proposed by \cite{Cen:03a}, or there may have been an
extended period of partial ionisation with a gradual transition to the
epoch observed in quasars.  If the IGM had a sudden transition from
neutral to completely ionised, the measured optical depth \taue$=0.17$
corresponds to a redshift of reionisation \zrei$=17$ for the best fit
$\Lambda$CDM cosmological model measured by WMAP \citep{Spergel:03}.

The large \taue measured by WMAP implies that an additional source of
ionising radiation is important at high-redshift . Two categories of
sources have been considered in previous studies: (i) stellar or AGN
sources, forming at $z \simlt 40$ in the first galaxies
\citep[\eg,][]{Cen:03b, WyitheL:03, Somerville:03, CiardiFW:03,
  Sokasian:03} or (ii) exotic (unknown) sources such as decaying
neutrinos \citep[\eg,][]{Sciama:82, HansenH:03}, super-heavy dark
matter decay at recombination \citep{BeanS:03} or primordial BHs,
either {\it via} evaporation \citep{Guedens:02} or {\it via} accretion
\citep{GnedinOR:95}. In this paper (paper~I) we study the early
reionisation of the IGM by the first stars (\pop3).  In two companion
papers \citep*[paper~IIa and paper~IIb,][]{RicottiO:03, RicottiOG:03} we
study the partial ionisation by a putative X-ray background produced
by mass accretion on seed BHs formed in the first galaxies.

Several groups have studied, using analytical calculations
\citep[\eg,][]{Uehara:96, Larson:98, Nakamura:99} or numerical
simulations \citep[\eg,][]{Abel:02, BrommCL:99}, star formation in a
metal-free gas. Most groups agree \citep[but see][ for a more
conservative point of view]{OmukaiP:03} that the first population of
stars was probably massive, with typical mass of about 100 M$_\odot$.
The stellar IMF in the early universe is unknown but it is expected to
be top-heavy or bimodal \citep{Nakamura:01}. It seems natural
therefore that at high-redshift the emissivity of ionising radiation
was larger than $\epsilon_{\rm UV} \approx 1.3 \times 10^{-4}$ found
for a Salpeter IMF.  If all stars were massive ($M_* \simgt 10$
M$_\odot$), the emission efficiency of ionising radiation would be a
factor of 10-20 larger than for the Salpeter IMF.  During their
lifetime massive stars convert a little less than half of their
hydrogen into helium. The efficiency of thermonuclear reactions for
the p-p chain is about 7 Mev per hydrogen atom. If we assume that all
the energy liberated is emitted in ionising photons we obtain a
theoretical upper limit for the UV emission efficiency $\epsilon_{\rm
  UV}^{\rm max} \simlt 2-3 \times 10^{-3}$ \citep[\cf,][]{Miralda:03}.  The
maximum emission efficiency $\epsilon_{\rm UV}^{\rm max}$ can be
obtained only if low-mass stars do not form and if all the ionising
radiation can escape from the molecular clouds and the diffuse ISM of
the host galaxy. Here \euv is the ratio of energy density of the
ionising radiation field to the gas rest-mass energy density converted
into stars ($\rho_* c^2$). Thus, $\epsilon_{\rm UV} =(\overline{h_p
  \nu} / m_Hc^2) N_{\rm UV}^*$, where $N_{\rm UV}^*$ is the emitted
number of ionising photons (with mean energy $\overline{h_p \nu}$) per
baryon converted into stars.  At first glance, the enhanced efficiency
of UV production from \pop3 stars seems to be the natural explanation
for the WMAP result. But as noted earlier there are several issues
that need to be addressed that make such a model difficult to achieve:
\bi
\item Complete reionisation at increasingly high redshift requires a
  larger number of ionising photons per baryon.  The recombination
  rate of hydrogen is large and the fraction of radiation escaping
  each galaxy is likely to be less than unity.
\item The efficiency of star formation in small-mass haloes ($M_{\rm
    dm}<10^8$ M$_\odot$) is regulated by radiative feedback effects
  (\ie, star formation is reduced as a result of gas heating by UV,
  winds and SNe produced by star formation) and, according to
  cosmological simulations \citep{RicottiGSa:02, RicottiGSb:02}, these
  galaxies cannot reionise the IGM. The simulations have shown that
  even assuming a top-heavy IMF, the star formation is strongly
  suppressed and the mean stellar mass in small-mass galaxies
  decreases.  The simulation results agree with observations of the
  smaller-mass galaxies (dwarf spheroidal galaxies) in the Local Group
  that show the same large mass-to-light ratios \citep{Ricotti_pr:03}.
\item Later forming galaxies with masses $M_{\rm dm} \simgt 10^9$
  M$_\odot$ have a larger star formation efficiency and their stars
  could reionise the IGM. But the enhanced energy injection by SN
  explosions due to a top-heavy IMF would also reduce star formation
  in larger galaxies.  If the energy of the SNe is normal
  ($E_{51}=E/[10^{51} {\rm erg}]=1$), the SN energy input for a
  top-heavy IMF would be larger than for a Salpeter IMF. Some massive
  stars are known to explode as hypernovae with energies $E_{51} \gg
  1$.  Supermassive stars with masses $140 {\rm M}_\odot<M_*< 260$
  M$_\odot$ are thought to end their lives with an explosion
  (pair-instability SN) that does not leave any remnant. The explosion
  energy of pair-instability SNe (PIS) is proportional to their mass
  and is typically $E_{51} \simgt 100$.
\item SN explosions eject metals into the ISM and IGM.  The transition
  from a top-heavy IMF to a Salpeter IMF is thought to depend on the
  gas metallicity.  If the metallicity is $Z \simgt 10^{-4} - 10^{-5}$
  Z$_\odot$ \citep{Omukai:00, BrommF:01, Schneider:02} the gas can
  cool more efficiently and to lower temperatures, allowing
  fragmentation and the formation of small-mass stars. The importance
  of \pop3 stars for reionisation depends on how long metal-free gas
  is present in protogalaxies.  This issue is difficult to address
  because it requires a good understanding of the mixing and transport
  processes of the heavy elements, but metals produced by SNe will
  clearly lead to an environment in which a top-heavy mass function no
  longer exists \citep[\eg, see][]{WadaV:03}.  The only ways to avoid
  contamination that we have thought of are two: (i) have such
  efficient ejection of metals from star forming regions that lower
  density regions are contaminated before higher density ones ($d \ln
  Z/Z_\odot/dln \langle \rho \rangle<0$) or (ii) have such inefficient
  local mixing that effective homogenisation is delayed for $0.5-1$
  Gyrs, $\sim 5-10$ local dynamical times. We will return to these
  possibilities later.  
\ei
  
This paper is organised as follows. In \S~\ref{sec:dim} we show, using
order of magnitude estimates, that a large \taue favours a scenario in
which a large fraction of the first massive stars collapse into
primordial massive BHs. In \S~\ref{sec:sim} we show the results of a
set of cosmological simulations with radiative transfer. We focus on
the effects of the limited mass resolution of the simulations and
feedback from SN explosions. In \S~\ref{sec:sem} we use a semianalytic
calculation to estimate the effect of cosmological parameters on
\taue. In \S~\ref{sec:con} we present our conclusions.

\section{BH-forming or pair-instability Supernovae?}\label{sec:dim}


In this section we use simple arguments to estimate the number of
ionising photons needed to reionise the IGM. Since, conventionally, the
massive stars that produce ionising radiation end their lives as SN
explosions and BHs, we estimate the number density of metals and relic
BHs expected given the redshift of reionisation (or \taue). We show
how the observed metallicity of the IGM and ISM in galaxies and the
demographics of supermassive black holes (SMBH) at $z=0$ can put some
loose constraints on the IMF of the first stars.  We find that it is
very difficult to reionise the universe at high redshift without
strongly contaminating it with metals. These calculations admittedly
rely on simplistic assumptions but are useful to show the problems
related with each scenario.

Assuming ionisation equilibrium we have,
\begin{equation}
{dn_{\rm ph} \over dt} \sim {n_{\rm ph} \over t_{\rm H}} \sim {\overline
  n_{\rm e} \over \langle t_{\rm rec}\rangle},
\label{eq:nph}
\end{equation}
where $n_{\rm ph}$ is the number density of ionising photons,
$\overline n_{\rm e}$ is the mean electron number density, $t_{\rm H}$
is the Hubble time and $\langle t_{\rm rec}\rangle=(\alpha_B n_e
C_{HII})^{-1}$ is the mean recombination time inside the \HII
regions. The recombination and the Hubble times are, respectively,
\begin{eqnarray}
\langle t_{\rm rec}\rangle &\sim& 1.7 {\rm Gyr} \left({\Omega_{\rm b} h^2 \over 0.02}\right)\left({1+z \over
  7}\right)^{-3}C_{\HII}^{-1},\\
t_{\rm H} &\sim& 1 {\rm Gyr} \left({1+z\over 7}\right)^{-1.5} \left({h \over 0.7}\right)^{-1},
\end{eqnarray}
where $C_{\HII}=\langle n_{\HII}^2 \rangle/\langle
n_{\HII}\rangle^2$ is the mean effective clumping factor of the ionised gas.  The
number of ionising photons emitted per hydrogen atom (also counting
the number of recombinations per Hubble time) sufficient to fully
reionise the IGM at redshift \zrei is given approximately by
\begin{equation}
N_{\rm ph}\equiv {n_{\rm ph} \over n_{\rm H}} \sim \max({t_H/\langle
  t_{\rm rec}\rangle },~1) \sim
\max([(1+z_{\rm rei})/14]^{1.5}C_{\HII},~1),
\label{eq:taue0}
\end{equation}
where $n_{\rm H}$ is the mean hydrogen density. Assuming sudden
reionisation at redshift \zrei$>(14C_{\HII}^{-2/3}-1)$ and
neglecting the redshift dependence of the clumping factor, we have
$N_{\rm ph} \propto (1+z_{\rm rei})^{1.5}$.  The
optical depth to Thomson scattering is given by
\begin{equation}
\tau_e= c \sigma_T \int dt n_e \sim 0.069 {\Omega_{\rm b} h \over
  \Omega_{0}^{1/2}}  \int dz (1+z)^{1/2} \langle x_e \rangle_{\rm M},
\label{eq:taue1}
\end{equation}
where $\langle x_e \rangle_{\rm M}$ is the mass weighted electron
fraction.  Assuming a sudden reionisation at redshift \zrei we have
\taue$\propto (1+z_{\rm rei})^{1.5}$, that has the same redshift
dependence of $N_{\rm ph}$. Therefore, in the case we neglect the redshift
dependence of the clumping factor in \eq~(\ref{eq:taue0}), we find
that $N_{\rm ph}$ is proportional to \taue. A similar relationship was found
by \cite{Oh:03} that, owing to the cancellation of the electron
density in \eq~(\ref{eq:taue1}), also find the following expression:
\begin{equation}
N_{\rm ph} ={\tau_e \over n_{\rm H} \langle
  t_{\rm rec} \rangle c \sigma_T} \sim 10 \left ({\tau_e \over 0.1}\right),
\label{eq:taue}
\end{equation}
that it is independent of redshift. The results of numerical
simulations, presented in \S~\ref{sec:res}, will show that this expression is
not very accurate. A better agreement can be obtained using the equation
\begin{equation}
N_{\rm ph} \sim 10 \left ({\tau_e \over 0.1}\right)^{4},
\label{eq:taue_new}
\end{equation}
that has been empirically derived from the simulations. The reason for
this result is that the mean effective clumping factor inside the \HII
regions it is not constant with redshift because it depends on the
spatial distribution of the sources that are responsible for
reionisation. The effective clumping factor, $C_{\HII}$, is larger
for reionisation at higher redshift because the sources, being more
numerous and with smaller mean luminosity, must ionise first a large
fraction of the dense filaments in which they are located. In the
following calculations we use \eq~(\ref{eq:taue}) because we want to
use the most conservative assumption on the number of ionising photons
$N_{\rm ph}$. Using \eq~(\ref{eq:taue_new}) would only change the exponent
of \taue from unity to four in the following relationships of this
section.

If the massive stars that produce ionising radiation end their lives
exploding as SNe and eject most of their mass enriched with heavy
elements, the metal yield, $Y$, of a star population
is approximatively proportional to the number of ionising photons
emitted, independently of the assumed IMF \citep[\eg,][]{Madau:96}.
Using the population synthesis model STARBURST \citep{Leitherer:99},
we found that a population of stars with metallicity $Z \simgt 0.05$
Z$_\odot$ (\popII) has a yield
\begin{equation}
Y=15 g~{\rm Z}_\odot \left({\epsilon_{\rm UV} \over 2 \times 10^{-3}}\right),
\label{eq:yield}
\end{equation}
where $0.3<g<2$ depends weakly on the metallicity and \euv is the
efficiency of ionising radiation emission from stars. As noted in the
introduction, the maximum efficiency for thermonuclear reactions is
$\epsilon_{\rm UV}^{\rm max} \simlt E_{\rm p-p}/m_{\rm H}c^2 \simlt
2-3 \times 10^{-3}$, where $E_{\rm p-p}$ is the mean energy per
nucleon emitted in form of ionising radiation. In this calculation we
have assumed that \pop3 stars during their lifetime are able to
convert half of their hydrogen into helium and that they are hot
enough to emit all their energy into ionising photons
\citep[\cf,][]{Miralda:03}.  The yield of \pop3 is very uncertain. The
main uncertainty comes from the unknown fraction of their stellar mass
and metals that implodes into a black hole. In the following paragraph
we will show that, if \pop3 stars have ``normal'' yields (neglecting
that a substantial fraction of their mass and metals is locked into
the black hole remnant) they cannot produce the large \taue measured
by WMAP without a production of metals that is not consistent with the
observed metallicities in the IGM and ISM.


The emissivity of ionising photons by massive stars is
\[
 \epsilon_{\rm UV} \simeq X (1.36 \times 10^{-8}) N_{\rm UV}^*,
\]
where $X=\overline{h\nu}/13.6 eV \approx 2-3$ is the mean energy of
\HI ionising photons and $N_{\rm UV}^*$ is the number of ionising
photons per baryon converted into stars.  If $f_*$ is the fraction of
baryons converted into stars we have $N_{\rm ph} \equiv N_{\rm
  UV}^*f_*$.

Let's assume that massive \pop3 stars have the same yields as massive
\popII stars.  Since the total metal production is $Z=Yf_*$, using
\eq~(\ref{eq:yield}), we have
\begin{eqnarray}
f_* &=& 6.8 \times 10^{-6}\left({2 \times 10^{-3} \over
  \epsilon_{\rm UV}}\right) X N_{\rm ph} ~~~{\rm
  and}\\
Z &=& (1g \times 10^{-4}~{\rm Z}_\odot) X N_{\rm ph}.
\end{eqnarray}
The fraction of baryons that needs to be converted into stars to get a
given \taue can be estimated using \eq~(\ref{eq:taue}) and using $X=3$:
\[
f_* \sim 2 \times 10^{-4} \left({2 \times 10^{-3} \over \epsilon_{\rm UV}}\right) \left({\tau_e \over 0.1}\right).
\]
This relationship shows that, if the IMF is top-heavy, a smaller
fraction of baryons needs to collapse into stars. We can also estimate
the mean star fraction of galaxies $f_*^{\rm gal} = f_*/f_{\rm coll}$,
where $f_{\rm coll}$ is the fraction of baryons in virialised DM
haloes. At $z=16$, most of the collapsed baryons are in small-mass
galaxies. The mass fraction of DM haloes with mass $M_{\rm dm} <10^8$
M$_\odot$ is $\Omega_{\rm dm} \sim 1.74$ \% and, $\Omega_{\rm dm} \sim
0.05$ \% in haloes with mass $M_{\rm dm} >10^8$ M$_\odot$. If we
assume that the baryons follow the DM ($f_{\rm coll}=\Omega_{\rm
  dm}$), the fraction of gas converted into stars must be
\begin{equation}
f_*^{\rm gal} \sim 
\begin{cases}
40 \% \left({2
  \times 10^{-3} \over\epsilon_{\rm UV}}\right)\left({\tau_e \over
  0.1}\right)~~\text{if $M_{\rm dm} >10^8$ M$_\odot$}\\
1 \% \left({2
  \times 10^{-3} \over\epsilon_{\rm UV}}\right)\left({\tau_e \over
  0.1}\right)~~\text{if $M_{\rm dm}<10^8$ M$_\odot$}.
\end{cases}
\end{equation}
If galaxies in small-mass haloes do not contribute to reionisation,
the efficiency of star formation in larger galaxies needs to be
approximately $30-50$ \%, which is perhaps larger than the efficiency
estimated for globular clusters. If small-mass galaxies contributed
significantly to reionisation, then their efficiency of gas conversion
into stars must be larger than $1-2$ \%.  Assuming a ``closed box''
chemical evolution model for the galaxy, the ISM metallicity is
\begin{equation}
Z_{\rm ISM} = Y f_*^{\rm gal}=
\begin{cases}
6g~{\rm Z}_\odot \left({\tau_e \over 0.1}\right)~~\text{if $M_{\rm dm}
  >10^8$ M$_\odot$}\\
0.15g~{\rm Z}_\odot \left({\tau_e \over 0.1}\right)~~\text{if $M_{\rm dm}
  <10^8$ M$_\odot$}.
\end{cases}
\end{equation}
These metallicities are far too large when compared to the
metallicities of the oldest star clusters known: old globular clusters
that typically have [m/H]=-1.5 \citep[\eg,][]{Vandenberg:96} and
dwarf spheroidals that have $-2.5<[m/H]<-1$ \citep[\eg,][]{Mateo:98}.
To match observations all but $10^{-2}$ to $10^{-3}$ of the metals
would need to have been ejected. With such a large {\it in situ} metal
production it is also difficult to have patches of metal-free gas so
that \pop3 stars can continue to form.
If instead we assume that most metals are ejected from the galaxy
haloes into the IGM, we find (assuming $X=3$) that the IGM mean
metallicity would be
\begin{equation}
Z_{\rm IGM}=3g \times 10^{-3}~{\rm Z}_\odot \left({\tau_e \over 0.1}\right).
\label{eq:zigm}
\end{equation}
But this value is about ten times larger than the estimated mean
metallicity $Z \simeq 2 \times 10^{-4}$ Z$_\odot$ of mildly overdense
regions in the IGM at redshift $z \simeq 3$ \citep{Schaye:03}. It
seems therefore that we need to find a mechanism to hide most heavy
elements. If metal enriched gas remains highly inhomogeneous in the
ISM, \pop3 stars could continue to form for a while, together with
normal stars. This scenario is not implausible but, because of the
formation of some small-mass stars, the efficiency of UV emission will
be reduced to $\epsilon_{\rm UV}<10^{-3}$ and reionisation would not
occur by this process.

Let's now relax the assumption that \pop3 stars have the same yields
as \popII stars. The initial mass function and metal yields from \pop3 stars
are extremely uncertain\footnote{.
  Metal-free stars in the mass range $10 {\rm M}_\odot< M_*< 130$
  M$_\odot$ have reduced yields because, being more compact than
  \popII stars, they are characterised by a large amount of heavy
  element fall-back onto the black hole remnant.  Stars more massive
  than $260$ M$_\odot$ do not explode as SNe but rather collapse
  directly into a BH without exploding. On the other hand, stars with
  masses $130 {\rm M}_\odot< M_* < 260$ M$_\odot$ provide the worst
  case for metal pollution and mechanical energy input. They explode
  as pair-instability SNe ($E_{51} \sim 10-100$), ejecting all of the
  metals without leaving any remnant and leaving a specific signature
  in the relative abundance distributions that has been sought for but
  not found \citep{VenkatesanT:03}.}  The simplest way to parametrise
these uncertainties is to assume that a fraction, $f_{\rm BH}$, of
\pop3 star mass and metals collapses onto BHs.  The baryon fraction in
seed BHs, $\omega_{\rm BH}=f_{\rm BH}f_*$, and the metal production
are given by
\begin{eqnarray}
\omega_{\rm BH} &\sim& f_{\rm BH} 2 \times 10^{-4}\left({2 \times
    10^{-3} \over \epsilon_{\rm UV}}\right)
\left({\tau_e \over 0.1}\right), \label{eq:bhl}\\
Z &\sim& (1-f_{\rm BH}) 3g \times 10^{-3}~{\rm Z}_\odot \left({\tau_e \over 0.1}\right). 
\label{eq:zl}
\end{eqnarray}
An upper limit for the total mass of seed BHs can be estimated
assuming that most of the mass in supermassive black holes (SMBHs)
observed in the bulges of galaxies at $z=0$ is built up from the
merger of seed BHs. We estimate a baryon fraction in SMBHs,
$\omega_{\rm BH, max}$, from the BH-bulge mass relation
\citep[\eg,][]{Kormendy:95} and from the estimated baryon fraction in
spheroids, $\omega_{\rm bulge}$, at $z=0$:
\[
\omega_{\rm BH, max}= \left({M_{\rm SMBH} \over M_{\rm bulge}} \right)
\omega_{\rm bulge} = (1 \pm 0.7) \times 10^{-4}. 
\]
We adopted the values $M_{\rm SMBH}/M_{\rm bulge}=(1.5 \pm 0.3)\times
10^{-3}$ \citep{Gebhardt:00} and $\omega_{\rm bulge}=(6.5 \pm 3)$\%
\citep{Persic:92, Fukugita:98}.  This is an upper limit because a
substantial fraction of the mass of SMBHs must be due to gas accretion
to account for the luminosity of quasars.  It is also possible,
however, that a fraction of seed BHs is expelled from their host
galaxies or does not end up in the observed population of SMBHs
\citep{MadauR:01}.  The recent discovery of several ultra-luminous
X-ray sources (ULX) has been interpreted as evidence for
intermediate-mass BHs (of about 100 M$_\odot$). These objects could be
relics of primordial BHs.  Theoretical studies have shown that the
expected luminosity function of accreting relic BHs in galaxies is
consistent with observations of ULX source
\citep{Islam:03a,Islam:03b}. Since we do not know how much they
contribute to the BH mass budget at $z=0$ in the calculation presented
here, we neglect their contribution. In any case, in paper~II and
paper~III we will show that the upper limit on $\omega_{\rm BH}$
adopted here is valid independently of the fate of the first BHs
because it is constrained by limits on the observed $\gamma$-ray
background.
If we use the baryon fraction in SMBHs at $z=0$, $\omega_{\rm BH,
  max}=10^{-4}$, as an upper limit for $\omega_{\rm BH}$ in \eq~(\ref{eq:bhl})
and $Z_{\rm IGM} = 2 \times 10^{-4}$ Z$_\odot$ as an upper limit for
the IGM metallicity given by $Z_{\rm IGM}=\langle f_{\rm ej}\rangle
Z$, where $\langle f_{\rm ej}\rangle$ is the fraction of metals
ejected into the IGM and $Z$ is from \eq~(\ref{eq:zl}), we find:
\[
1-{0.07 \over g \langle f_{\rm ej}\rangle} \left({\tau_e \over
    0.1}\right)^{-1}< f_{\rm BH} < 0.5 
\left({\epsilon_{\rm UV} \over 2 \times
10^{-3}}\right) \left({\tau_e \over 0.1}\right)^{-1}.
\]
We estimate that, if \pop3 stars reionised the IGM to the value
measured by WMAP (\taue$\sim 0.17$), the mass fraction of \pop3 stars
that might collapse into BHs is $f_{\rm BH} \sim 30$ \% and the metal
yield of \pop3 stars is less than half the ``normal'' value. This
result is rather extreme but plausible if \pop3 stars have masses
$M_*>260$ M$_\odot$ or if their explosion energy is $E_{51}<1$. Using
the same assumptions as before on the metal and BH budget, we find an
upper limit on the Thomson optical depth
\[
\tau_e < 0.05 \left({\epsilon_{\rm UV} \over 2 \times  10^{-3}}\right)
+ {0.007 \over g \langle f_{\rm ej}\rangle}.
\]

In conclusion, if the optical \taue$ \sim 0.17$ of the IGM is due to
ionising radiation from massive stars, we need to assume that most of
the heavy elements synthesised in the stars are not ejected into
either the ISM or IGM. Similar conclusions have been found in order to
prevent IGM over-enrichment with heavy elements by \pop3 stars, if
they have to account for the observed near-infrared background excess
after subtraction of `normal' galaxies contribution
\citep{Bond:86,Shchekinov:86,Santos:02,Salvaterra:03}.

A very inefficient mixing of the ejecta with primordial gas is also a
possible scenario. This would lead to the situation in which the high
density regions, preferred for subsequent star formation, are
contaminated by metals by more than the average amount but the
metals are not incorporated into subsequent stars formed locally until
after some delay. If the mechanism of metal ejection leads to metal
exclusion from the star-forming high density regions, multiple
generations of \pop3 stars could form in the same galaxy and in
neighbouring galaxies. In this scenario the gas metallicity and
overdensity are anticorrelated.  Numerical simulations fail to
reproduce this anticorrelation but this could be a consequence of the
insufficient numerical resolution and the badly understood physics of
metal transport. As a consequence most metals would reside in the \lya
forest at high redshift. Observation of the metallicity in the \lya
forest and the metallicity in quasar host galaxies at $z \sim 3$ do
not show anticorrelation of density and metallicity, but this scenario
cannot at present be ruled out at high redshift.
\begin{table*}
\centering
\caption{List of simulations with radiative transfer.\label{tab:1}}
\begin{tabular*}{14.5 cm}[]{ll|cccccccc}
\# & {RUN} & {$N_{\rm box}$} & {$L_{\rm box}$} & {Mass Res.} &
{Res. (com.)} & Smaller galaxy & 
{$\epsilon_*$}&{$\epsilon_{\rm UV}$\fesc} & $F_{\rm IMF}$ \\
{} & {} & {$h^{-1}$ Mpc} & {$h^{-1}$ M$_\odot$} & {$h^{-1}$ pc} & {$h^{-1}$ M$_\odot$} &
{} & {} & {}\\
\hline
1 & 64L1VM & 64 & 1.0  & $3.15\times 10^5$ & 976 & $3\times 10^7$ & 0.1 & 
$10^{-3}$ & 17\\
2 & 64L2VM & 64 & 2.0  & $2.52\times 10^6$ & 1952 & $2.5\times 10^8$ & 0.1 & 
$10^{-3}$ & 17\\
3 & 128L1VM & 128 & 1.0  & $3.94\times 10^4$ & 488 & $4\times 10^6$& 0.1 & 
$10^{-3}$ & 17\\
4 & 128L2VM & 128 & 2.0  & $3.15\times 10^5$ & 976 & $3\times 10^7$ & 0.1 & 
$10^{-3}$ & 17\\
5 & 128L4VM & 128 & 4.0  & $2.52\times 10^6$ & 1952 & $2.5\times 10^8$ & 0.1 &
$10^{-3}$ & 17\\
6 & 128L2VMb & 128 & 1.0  & $3.15\times 10^5$ & 976 & $3\times 10^7$ & 0.1 & 
$10^{-3}$ & 1\\
\end{tabular*}
\flushleft
{Parameter description. {\em Numerical parameters:}
   $N_{\rm box}^3$ is the number of grid cells and $L_{\rm box}$ is the box size
   in comoving \Mpc. The mass of
   the DM particles is listed in column four, the comoving
   spatial resolution in column five and the minimum DM mass of resolved
   galaxies (with at least 100 DM particles) is given in column six. 
{\em Physical parameters:} $\epsilon_*$ is the star
   formation efficiency, $\epsilon_{\rm UV}$ is the ratio of energy
   density of the ionising radiation field to the gas rest-mass energy
   density converted into stars (depends on the IMF), \fesc is the
   escape fraction of ionising photons from the resolution element and
   $F_{\rm IMF}$ is a weighting parameter for the energy input and
   metal yields from SN explosions, that differs from unity if
   the IMF is not a standard Salpeter IMF (see \S~\ref{sec:res}).}
\end{table*}
\begin{figure*}
\centerline{\psfig{figure=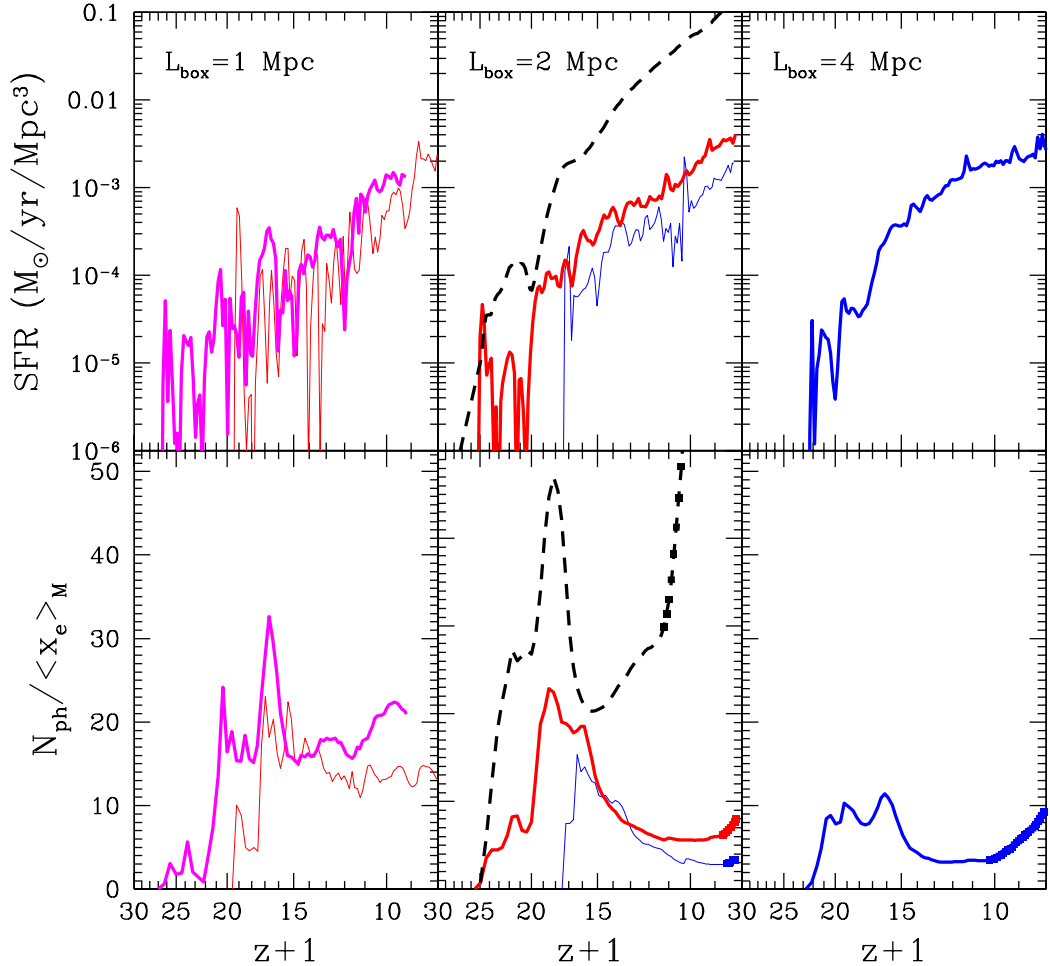,height=15cm}}
\caption{\label{fig:sfr} (top panels) Star formation rate as a function of
  redshift for the simulations in table~\ref{tab:1}. The thick lines
  refer to simulations with $N_{\rm box}=128$ and the thin lines refer
  to simulations with $N_{\rm box}=64$. All the simulations except
  128L2VMb (thick dashed line) include strong feedback from SN
  explosions and differ only in box size. The $L_{\rm box}=1$ \Mpc
  simulation resolves small-mass galaxies in which star formation is
  strongly suppressed by SN explosions. We achieve convergence in both
  $L_{\rm box}=2$ and $L_{\rm box}=4$ \Mpc simulations, where the
  galaxies that dominate the SFR at $z>10$ are numerically resolved.
  In simulation 128L2VMb we reduced the energy input from SN
  explosions by a factor of ten. The resulting SFR is about 20 times
  larger. Integrated number of ionising photons produced per free
  electron, $N_{\rm ph}/$\xe, as a function of redshift (bottom
  panels). The redshifts after reionisation are marked with points.}
\end{figure*}

Perhaps a more natural choice is to assume that most of the excess
metals are locked inside intermediate-mass BHs.  The recently
discovered lower metallicity star HE0107-5240 (with [Fe/H]$=-5.3$) is
iron deficient and carbon rich \citep{Christlieb:02}. The abundance
pattern of HE0107-5240 can be explained by assuming that the gas has
been enriched by a subluminous supernova explosion ($E_{51} \sim 0.3$)
of a zero-metallicity star with a mass of 25 M$_\odot$
\citep{UmedaN:03}. The iron metal yield is reduced because of
fall-back into the central BH. The abundance pattern of HE0107-5240 is
not compatible with yields from pair-instability SNe. To summarise,
even with extremely optimistic assumptions concerning the UV
efficiency of low metallicity stars we find that the only scenario
capable of ionising the universe at high redshift is one in which a
large fraction of the mass of \pop3 stars collapses onto massive black
holes.

\section{Cosmological Simulations}\label{sec:sim}

The results shown in the next section are based on numerical
cosmological simulations that include radiative transfer and feedback
from SN explosions. The code has been implemented to study stellar
reionisation and the formation of the first galaxies that is
self-regulated by radiative feedback effects in the early universe.
The simulations were run on COSMOS, an SGI origins 38000 in DAMPT,
University of Cambridge.

We adopt a concordance $\Lambda$CDM cosmological model with
parameters: $\Omega_0 = 0.3$, $\Omega_\Lambda= 0.7$, $h=0.7$ and
$\Omega_{\rm b} = 0.04$.  The initial spectrum of perturbations has
$\sigma_8=0.91$ and $n=1$.

The effective emissivity is $\epsilon_{\rm UV}^{\rm eff}=\langle
f_{\rm esc} \rangle \epsilon_{\rm UV}$. The value of \euv depends on
the initial mass function (IMF) and the metallicity of the stellar
population.  Assuming a Salpeter IMF with star masses between $1 {\rm
  M}_\odot \le {\rm M}_* \le 100 {\rm M}_\odot$, we have for \pop3
(\popII) stars $\epsilon_{\rm UV}=3 \times 10^{-4}(1.3 \times
10^{-4})$, $X=\overline{h_p \nu} / (13.6~{\rm eV}) =2.47(1.76)$
\citep[\eg,][]{RicottiGSa:02}. The quantity \fesc here is defined as
the mean fraction of ionising radiation that escapes from the
resolution element\footnote{Note that the definition of \fesc can
  create confusion. In numerical simulation, such as the ones
  presented in this work, \fesc is usually defined as subgrid physics,
  but in other works is defined as the mean fraction of ionising
  radiation escaping the galaxy haloes (\eg, the virial radii). The
  definition of \fesc and the definition of the clumping of the IGM
  are related and depend on how the outer edges of galaxies are
  defined.  It is therefore difficult to compare predictions of
  different works (numerical and analytical) and estimate which are
  reasonable values to assume for the free parameter \fesc.
  Nevertheless, \fesc cannot be treated purely as a freely adjustable
  parameter but has cosmological interest
  \protect{\citep[\eg,][]{Ricotti:03}}.}.  Assuming a Salpeter IMF,
\fesc must be in the range 20-50\% in order to have reionisation at
\zrei$\simeq 7$ \citep[\eg,][]{Benson:01, Gnedin:00}.

In this paper, as we are interested in determining an upper limit on
\taue, we adopt the maximum emissivity of ionising radiation
$\epsilon_{\rm UV}^{\rm max}$ from stellar sources produced by
thermonuclear reactions and \fesc$=0.5$.  Assuming a constant
emissivity is obviously not self-consistent since the metals ejected
by \pop3 stars eventually trigger the transition to the era when
\popII stars dominate the global star formation rate.  In our
simulations \pop3 stars are dominant for long enough to fully reionise
the IGM, but it should be kept in mind that, as noted above, due to
metal enrichment, a complete reionisation by \pop3 stars might never
be achieved.  In the next subsection we describe the main features of
the code.

\subsection{The Code\label{ssec:code}}
The simulations were performed with the ``Softened Lagrangian
Hydrodynamics'' (SLH-P$^3$M) code described in detail in
\cite*{Gnedin:95}. The cosmological simulation evolves collisionless
DM particles, gas, ``star-particles'', and the radiation field in four
frequency bins: optically thin radiation, \HI, \GI and \GII ionising
radiation fields. The radiative transfer is treated self-consistently
(\ie, coupled with the gas dynamics and star formation) using the
OTVET approximation \citep{GnedinA:01}. The star particles are formed
when the dense cooling gas in each resolution element sinks below the
numerical resolution of the code. The code adopts a deformable mesh to
achieve higher resolution in the dense filaments of the large scale
structure. We solve the line radiative transfer in the \H2
Lyman-Werner bands for the background radiation. We include secondary
ionisation of H and He, heating by Ly$\alpha$ scattering, detailed \H2
chemistry and cooling, and self-consistent stellar energy distribution
of the sources \citep{RicottiGSa:02}.

\subsubsection{Feedback from SN explosions}
We include the effects of SN explosions using the method discussed in
\cite{Gnedin:98a}. The current resolution of cosmological simulations
does not allow numerical resolution of radiative shock fronts and the
modelling of SN explosions from ``first principles''. A semianalytic
recipe has to be included at the sub-grid level. This treatment
introduces new free parameters in the simulation and the results might
depend on the model implemented. The approximation used to solve
radiative transfer is instead much more reliable and has been shown,
for simple test cases, to reproduce the exact solution.  This is part
of the reason why in our previous study on the formation of the first
galaxies \citep{RicottiGSb:02} the effect of SN explosions was not
included. In the present paper we find that, assuming a Salpeter IMF,
the effect of SN explosions is not very important, so that omission in
the earlier is justified. The main conclusions in \cite{RicottiGSb:02}
remain unchanged if we include SN feedback and the radiative feedback
is strong enough (\ie, assuming a Salpeter IMF and \fesc$\simgt
10$\%). However, in the current work assuming a top-heavy IMF, the SN
energy injection has more dramatic effects and also affects galaxies
with masses $M_* \simgt 10^9$ M$_\odot$.

\subsection{Results}\label{sec:res}

In table~\ref{tab:1} we list the parameters adopted in the
simulations.  Here $\epsilon_*$ is the star formation efficiency of
the adopted star formation law \citep{CenO:92}: $d\rho_*/dt =
\epsilon_* \rho_{\rm g} / t_*$, where $\rho_*$ and $\rho_{\rm g}$ are
the stellar and gas density, respectively. The quantity $t_*$ is the
maximum of the local dynamical and the local cooling time. The escape
fraction from a cell, \fesc, is resolution-dependent. The parameter
$F_{\rm IMF}$ is proportional to the mean metallicity yields and
energy input by SN explosions of the stellar populations.  This
parameter is unity for a Salpeter IMF and \popII stars.  But $F_{\rm
  IMF}>1$ if the IMF is top-heavy and pair-instability SNe are
dominant. Population~III stars could have $F_{\rm IMF}<1$ if
super-massive stars, that collapse directly into BHs without exploding
as SNe, dominate the mass function.

\subsubsection{Convergence study: \fesc and photoevaporation of minihaloes}

\begin{figure}
\centerline{\psfig{figure=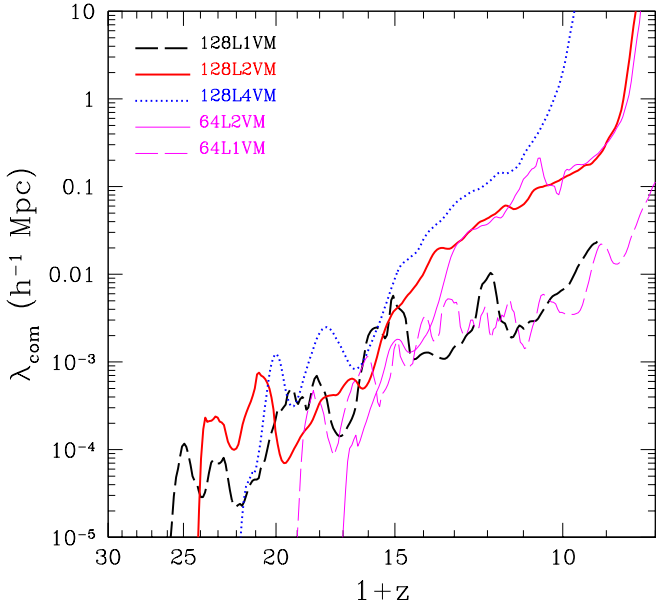,height=9cm}}
\caption{\label{fig:com} Comoving mean free path $\lambda_{\rm com}$ 
  of \HI ionising photons as a function of redshift. The size of \HII
  regions before overlap is approximately equal to $\lambda_{\rm
    com}$. Here we show $\lambda_{\rm com}$ for several simulations in
  table~\ref{tab:1} to point out the effects of changing the box size and
  mass resolution. }
\end{figure}
\begin{figure}
\centerline{\psfig{figure=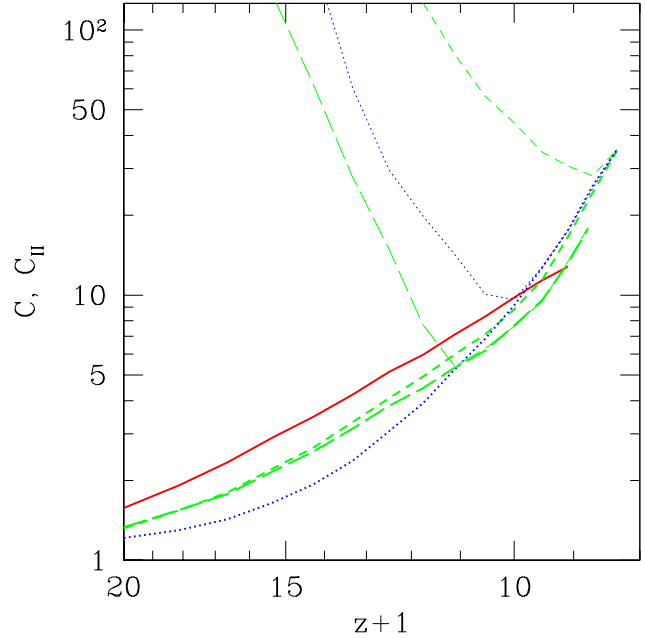,height=9cm}}
\caption{\label{fig:clump} Clumping factor of the IGM, $C$ (thick lines), and inside
  \HII regions, $C_{\HII}$ (thin lines), as a function of redshift.
  The solid, dashed and dotted lines show the clumping factor, $C$,
  for the strong SN feedback simulations with $N_{\rm box}=128$ and
  $L_{\rm box}=1, 2$ and 4 \Mpc, respectively, listed in
  table~\ref{tab:1}. The long-dashed lines refer to the $L_{\rm
    box}=2$ \Mpc simulation with weak SN feedback.}
\end{figure}
\begin{figure}
\centerline{\psfig{figure=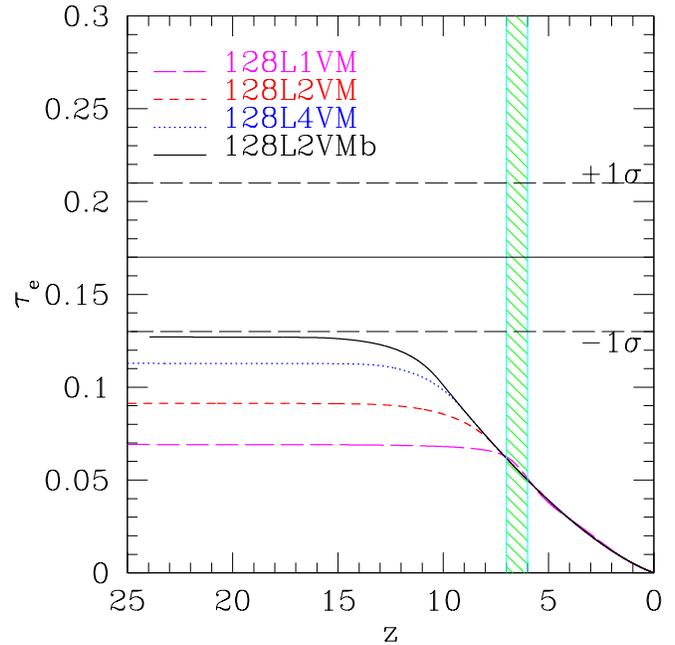,height=9cm}}
\caption{\label{fig:taue} Optical depth to Thomson scattering, \taue, as a
  function of redshift for the simulations with $N_{\rm box}=128$
  listed in table~\ref{tab:1}. The horizontal lines show the best
  value of \taue and $1\sigma$ confidence limits from WMAP
  \protect{\citep{Kogut:03}}. The vertical band indicates the redshift probed
  by the furthest quasars found by the SLOAN survey.}
\end{figure}

We have run a set of simulations with box sizes $L_{\rm box}=1, 2$ and
4 \Mpc and number of particles $N_{\rm box}^3=64^3$ and $N_{\rm
  box}^3=128^3$.  The other parameters of the simulations, listed in
table~\ref{tab:1}, are the same: we use the maximum efficiency of UV
production allowed by thermonuclear reactions, $\epsilon_{\rm UV}^{\rm
  max}=2 \times 10^{-3}$, and \fesc$=0.5$. We include the effect of SN
explosions with an energy input and metal production $17$ times the
value for a Salpeter IMF.  This is a reasonable assumption for a
top-heavy IMF, if pair-instability SNe are important.  But, given the
uncertainties in the energy of SN explosions and on the IMF of \pop3
stars, a scenario is also plausible in which \pop3 stars collapse
directly into BHs or produce subluminous SN explosions ejecting few
metals.  We explore this second scenario running a simulation with box
size $L_{\rm box}=2$ (128L2VMb in table~\ref{tab:1}) with reduced
energy input and metal ejection ($F_{\rm IMF}=1$).

In the top panels of \fig~\ref{fig:sfr} we show the global star
formation rate (SFR) as a function of redshift. In each panel, from
left to right, we show simulations with box size $L_{\rm box}=1, 2$
and 4 \Mpc.  The thick lines show the $N_{\rm box}=128$ simulations
and the thin lines the $N_{\rm box}=64$ simulations. The simulation
with weak feedback is shown in the central panel with a dashed thick
line. In the high-resolution $L_{\rm box}=1$ \Mpc simulation we
resolve star formation in galaxies with masses $M_{\rm dm} \simgt
3.94\times 10^6$ \Ms \citep[see][]{RicottiGSa:02}.  But, because of
the strong feedback from SN explosions, star formation in the
small-mass galaxies is suppressed.  This can be seen comparing the top
left panel to the simulations with larger box sizes and lower mass
resolution. In the larger box simulations the first stars form later
but the SFR quickly converges to the same value. The SFRs in the
$L_{\rm box}=2$ and $L_{\rm box}=4$ \Mpc box are almost identical. At
$z \sim 7$ the global SFR is $3 \times 10^{-3}$ M$_\odot$yr\m \Mpc
\mmm.  This value is still very small when compared to SFR$ \sim 0.1$
M$_\odot$yr\m \Mpc \mmm, estimated from observations at $z \sim 4-5$
\citep{Lanzetta:02}.

In the bottom panels of \fig~\ref{fig:sfr} we show the integrated
number of ionising photons per free electron, $N_{\rm ph}/$\xe, as a
function of redshift. The redshifts after reionisation (\ie,
\xe$>0.9$) are marked with points. The simulations shown are the same
as in the top panels.  At the redshift of reionisation the quantity
$N_{\rm ph}$ measures the number of recombinations per baryon. If
recombinations are negligible, $N_{\rm ph}$ is unity.  From the plots
it appears that the recombinations in the $N_{\rm box}=64$ simulations
(thin lines) are always underestimated with respect to the simulations
with higher mass resolution $N_{\rm box}=128$ (thick lines).  We see
that in the smaller scale boxes, which have higher mass resolution,
and hence allow for lower mass halo formation, the integrated number
of ionising photons produced per free electron required to reionise
the IGM is larger. This number initially increases as the Str\"omgren
spheres are finding their way through the denser regions surrounding
the ionising sources. When the volume filling factor of the \HII
regions is larger, the ionisation fronts propagate mostly in the
underdense regions and the integrated mean number of recombinations
per baryon, $N_{\rm ph}/$\xe, decreases to less than 10. In the left
panel the \HII regions never break out of the dense filaments, because
of the bursting mode of star formation in small-mass galaxies and the
absence of large ones in the small box. Because of the self-limiting
nature of the stellar feedback, as already found by
\cite{RicottiGSb:02}, small-mass galaxies are not able to reionise the
IGM. By comparing the middle and right panels we see that the
integrated mean number of recombinations per baryon at the redshift of
reionisation (marked by the first dot) is lower for the larger box
size, that has lower mass resolution.  This number is the same for
simulations that have the same mass resolution (\eg, the thin line in
the middle panel and the thick line in the right panel).
\begin{figure}
\centerline{\psfig{figure=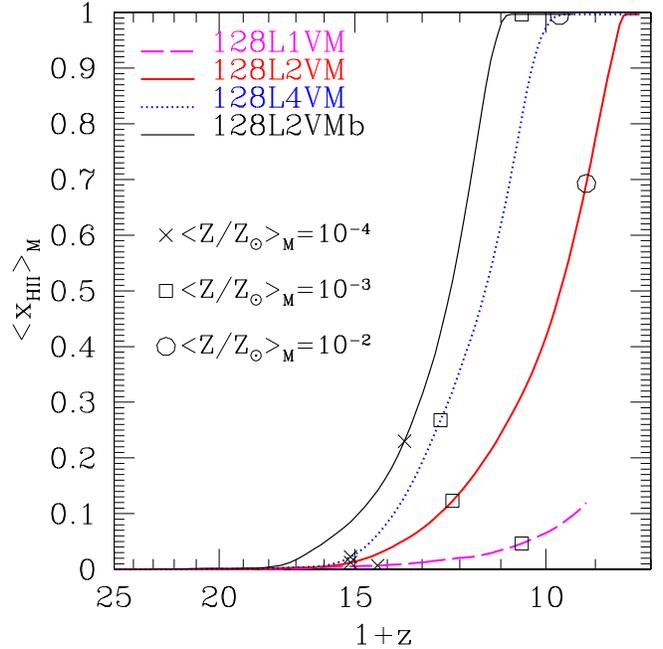,height=9cm}}
\caption{\label{fig:met} Hydrogen ionisation fraction as a
  function of redshift for the simulations with $N_{\rm box}=128$
  listed in table~\ref{tab:1}. The circles, squares and crosses show
  the redshift at which the mass weighted mean metallicity in each
  simulation is $Z=10^{-2}, 10^{-3}$, and $10^{-4}$ Z$_\odot$,
  respectively.}
\end{figure}

We have shown that the integrated number of recombinations at the
redshift of reionisation depends on the mass resolution of the
simulation. This means that low resolution simulations might
underestimate the clumping factor inside the \HII regions (\cf,
\fig~\ref{fig:clump}) or overestimate the mean value of \fesc.
Resolving small-mass haloes, that are photoevaporated during cosmic
reionisation consuming a fraction of the ionising photons, may also be
important to achieve a convergent value of \taue \citep{Shapiro:00,
  Haiman:01, Barkana:02}.  We also note that the bursting mode of star
formation in small-mass galaxies increases the number of
recombinations and thus increases the number of ionising photons
needed to reionise the IGM.
\begin{figure*}
\centerline{\psfig{figure=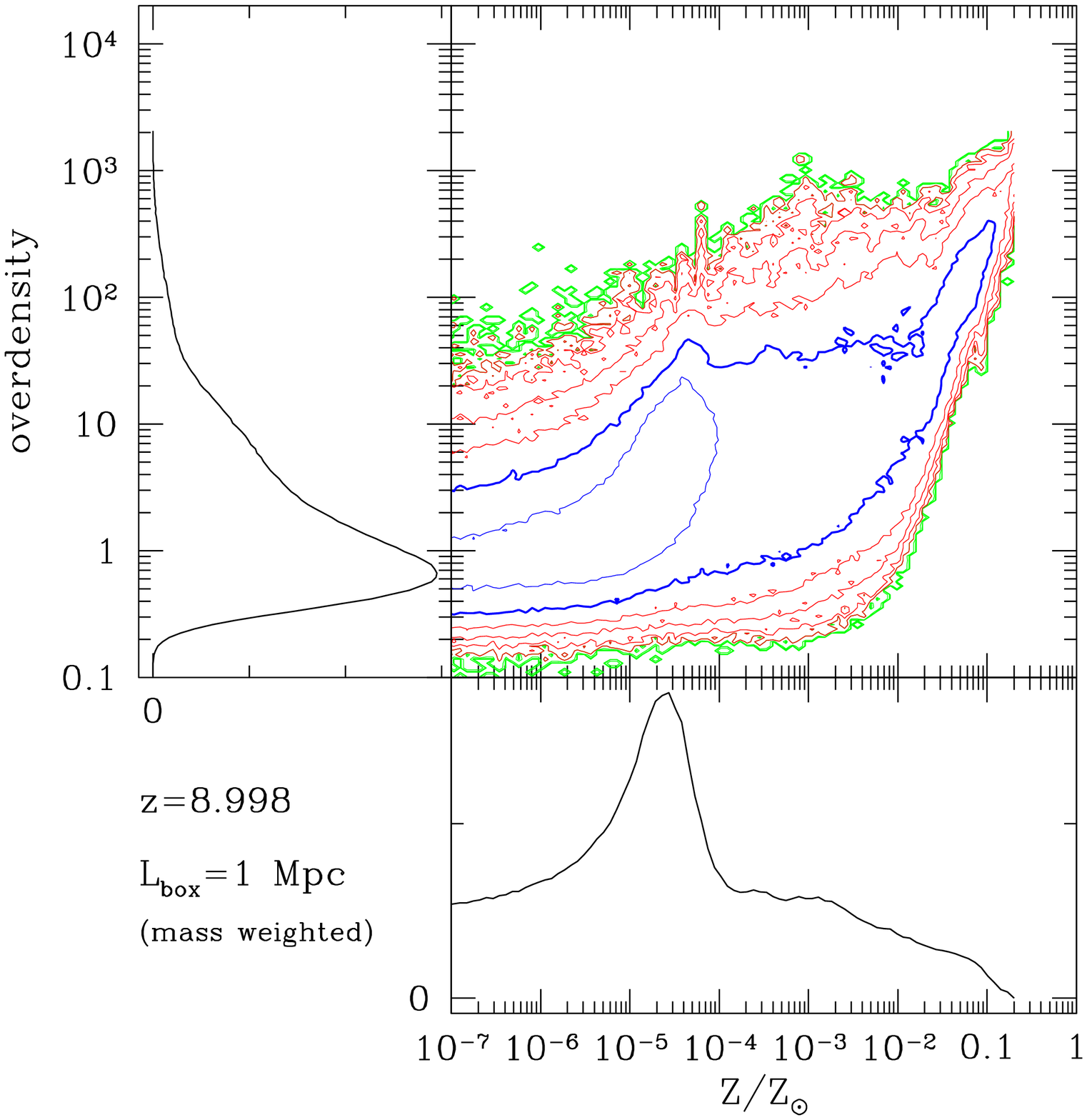,height=9cm}
\psfig{figure=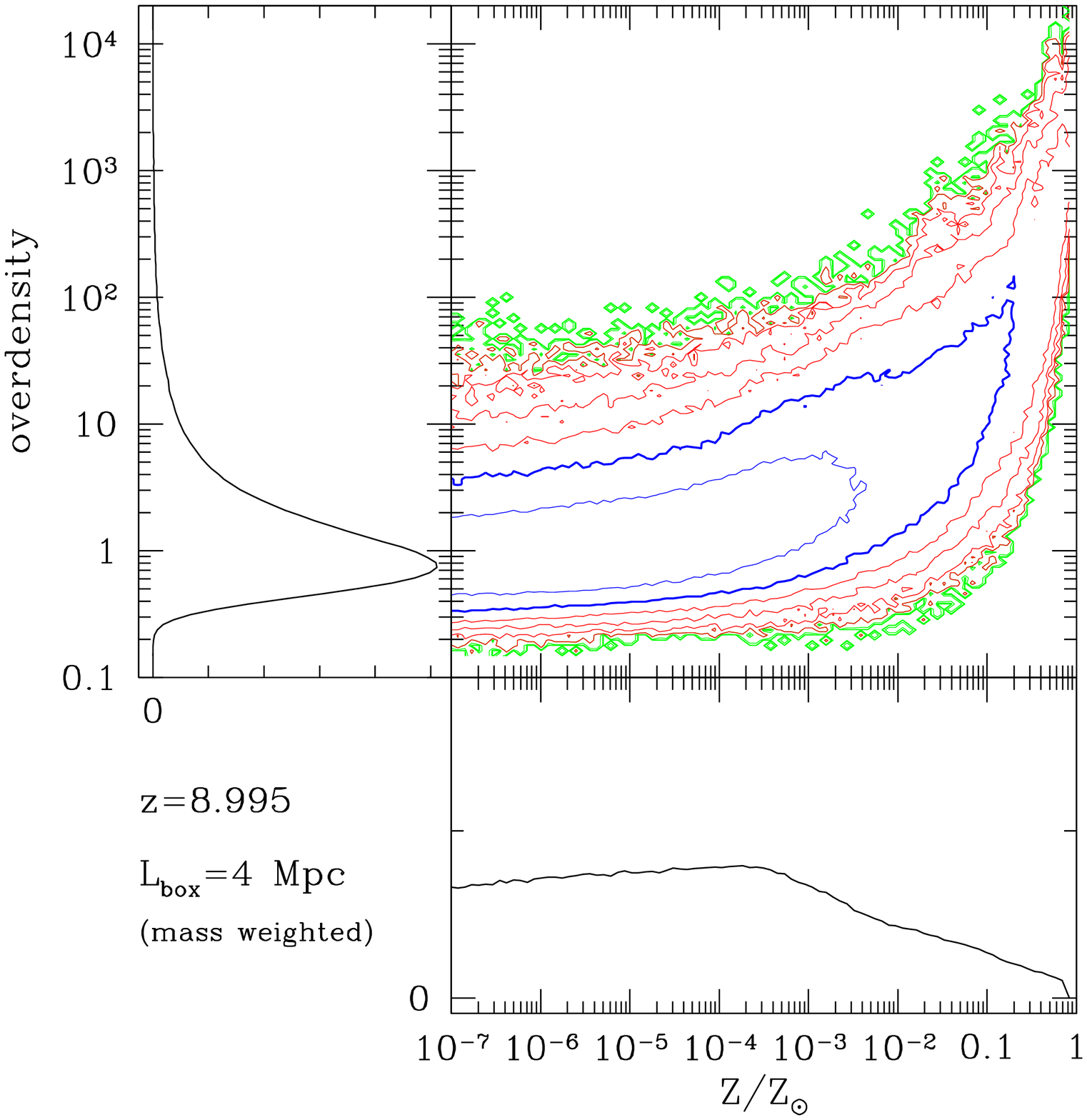,height=9cm}}
\caption{\label{fig:metdist} (left) Mass weighted
  metallicity-overdensity distribution for the simulation 128L1VM at
  redshift $z=9$. The presence of metal-enriched wind with $Z\sim
  10^{-4} Z_\odot$, produced by SN explosions in small-mass galaxies
  ($M>4 \times 10^{6}$ M$_\odot$), is evident in both the metallicity
  distribution and density distribution. (right) The same as in the
  left panel but for the simulation 128L4VM. The galaxies resolved in
  this simulation are less affected by SN explosions because of their
  larger masses ($M>2.5 \times 10^{8}$ M$_\odot$).}
\end{figure*}

In \fig~\ref{fig:com} we show the comoving mean free path
$\lambda_{\rm com}$ of \HI ionising photons as a function of redshift
for the simulations in table~\ref{tab:1} with strong feedback from SN
explosions. The simulations with $L_{\rm box}=2$ \Mpc and $L_{\rm
  box}=4$ \Mpc have approximately the same SFR but in the larger box
reionisation (defined as the redshift at which $\lambda_{\rm com}$
increases steeply) happens much earlier: \zrei$ \approx 7.5$ in the
$L=2$ \Mpc run and \zrei$\approx 9.5$ in the $L_{\rm box}=4$ \Mpc run.
The two $L_{\rm box}=2$ \Mpc simulations with different mass
resolutions ($N_{\rm box}=64$ and $128$) reionise the IGM at the same
redshift even if the SFR is half in the smaller resolution simulation.
This effect can also be seen comparing the $L_{\rm box}=1$ \Mpc
simulations with the others.  These results show that, even if we do
not need to resolve small-mass haloes to have a convergent global SFR,
we do need a high mass resolution or otherwise the optical depth to
electron Thomson scattering will be overestimated due to neglect of
recombination in high density regions.

\subsubsection{Strong and weak SN feedback}

In \fig~\ref{fig:taue} we show the optical depth to Thomson
scattering, \taue, as a function of redshift for the simulations with
$N_{\rm box}=128$ in table~\ref{tab:1}.  In \fig~\ref{fig:met} we show
the hydrogen ionisation fraction as a function of redshift for the
same simulations.  If we define \zrei as the redshift at which the
hydrogen ionisation fraction is 90\%, we find \zrei$=7.5$ for the
$L=2$ \Mpc simulation with strong SN feedback and \zrei$=10$ for the
simulation with weak SN feedback. For the same simulations the optical
depths are \taue$=0.09$ and \taue$=0.13$, respectively.  These
optical depths are considerably larger than expected for a sudden
reionisation at \zrei.  This is because the duration of reionisation,
from the redshift where the mass-weighted mean hydrogen fraction is 10
\%, to complete reionisation, is quite long. We find approximately
$\Delta z_{\rm rei}/ z_{\rm rei} \sim 50$ \%.  The maximum
\taue$\simlt 0.13$ found for the $L_{\rm box}=2$ \Mpc simulation is
obtained for the case when SN feedback is not important. But this
condition is not sufficient.  The maximum optical depth, \taue$\simeq
0.13$, can be achieved only if the metallicity of the gas in star
forming regions remains lower than a critical value. In the next
section we show that this requirement puts strong constraints on the
yield that \pop3 stars may have in order to contribute to the IGM
reionisation.

In summary, we do not achieve convergence for the simulations with
strong feedback.  Rare massive galaxies in this case contribute to the
global SFR, because star formation in the smaller ones is suppressed
by SN explosions. For this reason we need a box size of at least
$L_{\rm box}=4$ \Mpc to achieve a converged estimate for the SFR.  But
when we increase the box size the mass resolution becomes too small
and we underestimate the clumping of the IGM (\ie, the number of
recombinations). For this reason, even if the SFR has converged, \taue
is likely to be overestimated because of the too low recombination
rate in the $L_{\rm box}=4$ \Mpc box. The $L_{\rm box}=2$ \Mpc run
instead gives a better estimate of \taue because the underestimated
clumping of the IGM is compensated by the slightly underestimated SFR.
This can be seen by comparing the two $L_{\rm box}=2$ \Mpc runs with
$64^3$ and $128^3$ particles in the middle panels of
\fig~\ref{fig:sfr}: in the low resolution run the number of
recombinations (bottom panel) and the SFR rate are underestimated (top
panel). Fortunately these two effects almost cancel each other and,
even if the low resolution simulation (thin lines) is not convergent,
we did get an estimate of the redshift of reionisation (marked by the
last dot in the bottom panel) and \taue that is about the same as in
the higher resolution simulation (thick lines).

For the weak feedback case we should be close to convergence since we
can safely use the smaller box. In this case the SFR is large enough
that the contribution from rare massive galaxies is negligible.
Unresolved small-mass galaxies do not contribute substantially to the
SFR.  Ideally, we need larger mass resolution to be sure that we are
not underestimating recombinations. Note that \fesc is defined as the
fraction of ionising photons that escapes from the resolution element.
Therefore, by definition, it is resolution dependent and as we
increase the resolution of the simulations its value should approach
unity. We would use \fesc$=1$ if we could perfectly resolve galaxies
down to the star forming regions in the ISM. Note that by definition,
the effective clumping factor of the IGM is complementary to \fesc, in
the sense that it accounts for all the recombinations produced in
regions larger or equal to the resolution element.

\subsubsection{Metal enrichment}

\begin{figure}
\centerline{\psfig{figure=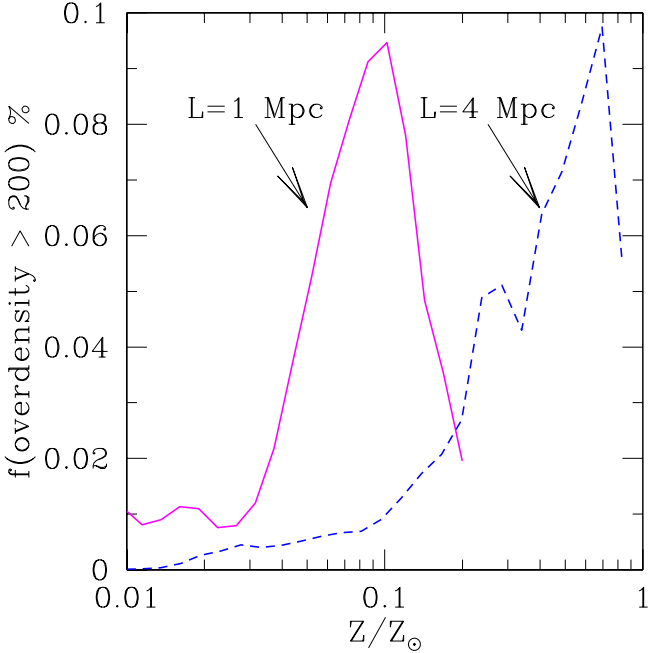,height=9cm}}
\caption{\label{fig:metd} Metallicity distribution of the gas mass
  fraction with overdensity larger than 200 at redshift $z=9$. The
  solid line refers to the simulation with box size $L_{\rm box}=1$
  \Mpc and the dashed line to the simulation with box size $L_{\rm
    box}=4$ \Mpc.}
\end{figure}
The circles, squares and crosses in \fig~\ref{fig:met} show the
redshift at which the mass weighted mean metallicity in each
simulation is $Z=10^{-2}, 10^{-3}$, and $10^{-4}$ Z$_\odot$,
respectively. The volume weighted mean metallicities (not shown here)
are ten times smaller than the mass weighted mean for both the
simulations with strong and weak SN feedback.  This means that most
metals are located in overdense regions.  This can be better seen in
\fig~\ref{fig:metdist}, where we show the mass weighted metallicity
distribution as a function of the overdensity for the simulation
128L1VM with $L_{\rm box}=1$ \Mpc (left) and 128L4VM $L_{\rm box}=4$
\Mpc (right). In both simulations the mean metallicity increases with
the overdensity. There is some scatter around the mean but at
overdensities larger than $200$ most of the gas has metallicity $Z\sim
0.1-0.01$ Z$_\odot$ (\cf, \fig~\ref{fig:metd}). These values are about
ten times larger than the mass-weighted mean metallicities at the same
redshift (or a hundred times the volume-weighted mean metallicities).
Similar results were found by \cite{CenO:99}. If $Z_{\rm cr} \simeq
10^{-4}-10^{-5}$ is the critical metallicity at which \popII stars
dominate over \pop3 stars, it is quite clear from \fig~\ref{fig:met}
that reionisation by \pop3 stars cannot be completed in any of our
simulations. Since stars are formed in overdense regions where the
metallicities are larger, \pop3 stars can not be important for
reionisation unless their yield is much smaller than assumed here or
the metals do not mix efficiently with metal-free gas. Note that we
require a factor of $10^3$ reduction in either yield or mixing
efficiency to keep the metallicity below the level required for
production of \pop3 stars (\cf, \fig~\ref{fig:metd}). While
\fig~\ref{fig:metd} implies that \pop3 is probably not relevant to
reionisation it is not intrinsically implausible or inconsistent with
observations. The quasar broad line emission regions sample the high
density gas in galaxies and the Sloan high redshift quasars ($z \sim
6$) already show a metallicity approaching solar values. This is
consistent with the most relevant ($L_{\rm box}=4$ \Mpc) curve in
\fig~\ref{fig:metd} and would remain so even if we allow for 50 \%
fall-back of metals onto black holes.
\begin{table*}
\centering
\caption{Results of simulations with radiative transfer.\label{tab:2}}
\begin{tabular*}{16.5 cm}[]{ll|cccccccc}
\# & {RUN} & {$\log f_*$} & {$\log f_*$} & {\taue} &
{\zrei} & $N_{\rm ph}$ & 
& {$\log \langle Z/Z_\odot \rangle$} at $z=9$ & \\
& & {at $z=15$} & {at $z=10$} & {} & {$\langle x_e \rangle_M=10^{-3}$} & {} & {volume weighted} & {mass weighted} & {$1+\delta>200$}\\
\hline
3 & 128L1VM  & $-5.70$& $-5.00$& 0.07 &  -  &  -  & $-4.00$ & $-2.70$ & -1.17 \\
4 & 128L2VM  & $-5.70$& $-4.70$& 0.09 & 7.5 & 6.5 & $-3.52$ & $-2.22$ & -0.66 \\
5 & 128L4VM  & $-5.70$& $-4.53$& 0.11 & 8.7 & 3.4 & $-3.40$ & $-2.10$ & -0.38 \\
6 & 128L2VMb & $-4.70$& $-3.22$& 0.13 & 9.9 & 30  & $-4.00$ & $-2.70$ & -1.11 \\
\end{tabular*}
\end{table*}

In the simulation with $L_{\rm box}=1$ \Mpc, the presence of
metal-enriched galactic winds with $Z\sim 10^{-4}$ Z$_\odot$, produced
by SN explosions in small-mass galaxies ($M>4 \times 10^{6}$
M$_\odot$), is evident in both the perturbed metallicity distribution
and density distribution (\cf, \fig~\ref{fig:metdist}). In the $L_{\rm
  box}=4$ \Mpc simulation galactic outflows are not important.  The
galaxies resolved in this simulation, because of their larger masses
($M_{\rm dm}>2.5 \times 10^{8}$ M$_\odot$), are less affected by SN
explosions and feedback.

In \tab~\ref{tab:2} we give the values of relevant quantities measured
in our high resolution simulations with parameters listed in
table~\ref{tab:1}. Here $\omega_*$ is the fraction of baryons
in stars, \zrei is the Gunn-Peterson reionisation redshift and $N_{\rm
  ph}$ is the number of ionising photons per baryon emitted at the
redshift of reionisation. We show the mean mass weighted and volume weighted
metallicities. In the last column we list the mean mass weighted
metallicity in regions with overdensities $>200$.

As noted in \S~\ref{sec:int}, the simple estimate presented in
that section for the number of ionising photons
$N_{\rm ph}$ as a function of the optical depth \taue,
\begin{equation}
N_{\rm ph} \sim 10 \left ({\tau_e \over 0.1}\right)^\alpha,
\end{equation}
with $\alpha=1$ (\cf, \eq~[\ref{eq:taue}]), is not very accurate when
compared to the simulations. A better agreement with the simulations
can be obtained using instead $\alpha=4$.  The reason for this result
is that the mean effective clumping factor inside the \HII regions it
is not constant with redshift as was initially assumed.  Using
$\alpha=4$ and $g=2$ in \eq~(\ref{eq:zigm}), the estimated IGM
metallicity as a function of \taue is in good agreement with the
simulation results listed in \tab~\ref{tab:2}. Since in
\eq~(\ref{eq:zigm}) we have assumed that all the metals are ejected
from the galactic haloes we should compare \taue (given in the third
column) to the mass weighted metallicities (given in the seventh
column).

\section{Semianalytic simulations}\label{sec:sem}

In this section we use a simple semianalytic model to study the
reionisation history of the IGM. This model can be useful to study the
dependence of reionisation history on the assumed cosmological model
and parameters. The semianalytic model do not include the effects of
radiative feedback, SN feedbacks and metal enrichment. It also uses
the results of the numerical simulations on the clumping factor of the
IGM. Given these assumptions and limitations (due to neglecting
feedbacks) the results of the semianalytic models are in good
agreement with the numerical simulations. For example, the value of
\taue$=0.13$ found in our numerical simulation with weak SN feedback
(128L1VMb) is shown as a circle in \fig~\ref{fig:pk}. The
corresponding value found in the semianalytic model, shown with a
square, is about $15$\% larger.

\subsection{The code}
We implemented a semianalytic model to study reionisation, chemical
evolution and re-heating of the IGM.  The method to calculate the
filling factor of \HII regions and the SFR is the same as in
\cite{Chiu:00}. But we also include X-ray ionisation of the IGM before
complete overlap of the \HII regions. Since in this paper we neglect
the X-ray background, a full description of the code will be given in
Appendix~A of paper~II. Here we only briefly summarise the main
features of the semianalytic model. The mass function of DM haloes and
their formation/merger rates are calculated using the extended
Press-Schechter formalism.  The SFR is assumed to be proportional to
the formation rate of haloes.  The effects of cooling and dynamical
biases that depend on the mass of the haloes are taken into account.
We consider the IGM as a two-phase medium: one phase is the ionised
gas inside the \HII regions and the other is the neutral or partially
ionised gas outside. We solve the radiative transfer for the volume
averaged specific intensity and we derive the specific intensity
inside and outside the \HII regions from their volume filling factor
and separating the contribution from local UV sources and the
background radiation from distant sources.  We solve the
time-dependent chemical network for eight ions (H, H$^+$, H$^-$,
H$_2$, H$_2^+$, He, He$^+$, He$^{++}$), and the thermal evolution as a
function of the gas density outside the \HII regions. Given the
clumping factor of fully ionised gas around the UV sources (taken from
the SLH simulations) we evolve the filling factor, temperature and
chemistry inside the \HII regions.  The following heating and cooling
processes are included: collisional- and photo-ionisation heating,
ionisation by secondary electrons, H, He, \H2 cooling, and Compton and
adiabatic cosmological cooling. The rates are the same as in
\citep{RicottiGS:01}.

\subsection{Dependence of \taue on small-scales power}
\begin{figure}
\centerline{\psfig{figure=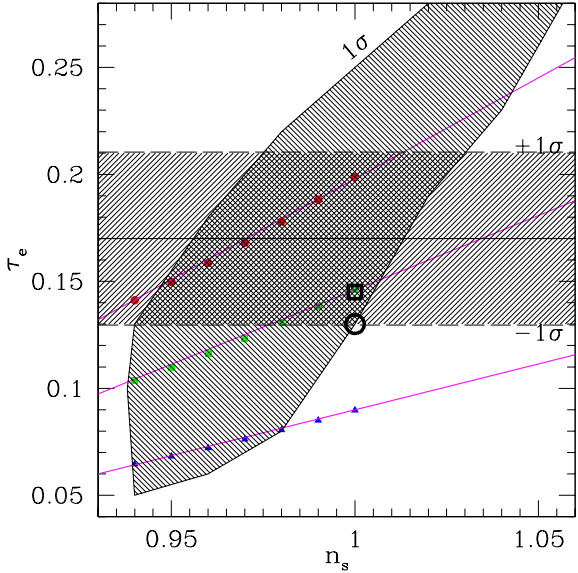,height=8cm}}
\caption{\label{fig:pk} Optical depth to Thompson scattering, \taue, 
  as a function of the spectral index of initial density
  perturbations, $n_s$, computed from the semianalytic code.  Here
  \sig8 is calculated according to the amplitude of perturbations at
  $k=0.05$ h Mpc$^{-1}$ measured by WMAP. Each line from top to bottom
  shows \taue$(n_s)$ for effective efficiencies of UV emission
  $\epsilon_{\rm UV}^{\rm eff}=10^{-2}, 10^{-3}$ and $10^{-4}$,
  respectively. We show the $1\sigma$ confidence contour of
  \taue-$n_s$ \protect{\citep{Spergel:03}} and of \taue
  \protect{\citep{Kogut:03}}. The value of \taue$=0.13$ found in our
  numerical simulation with weak SN feedback (128L1VMb) is shown as a
  circle. The corresponding value found in the semianalytic model,
  shown with a square, is about $15$\% larger.}
\end{figure}
\begin{figure}
\centerline{\psfig{figure=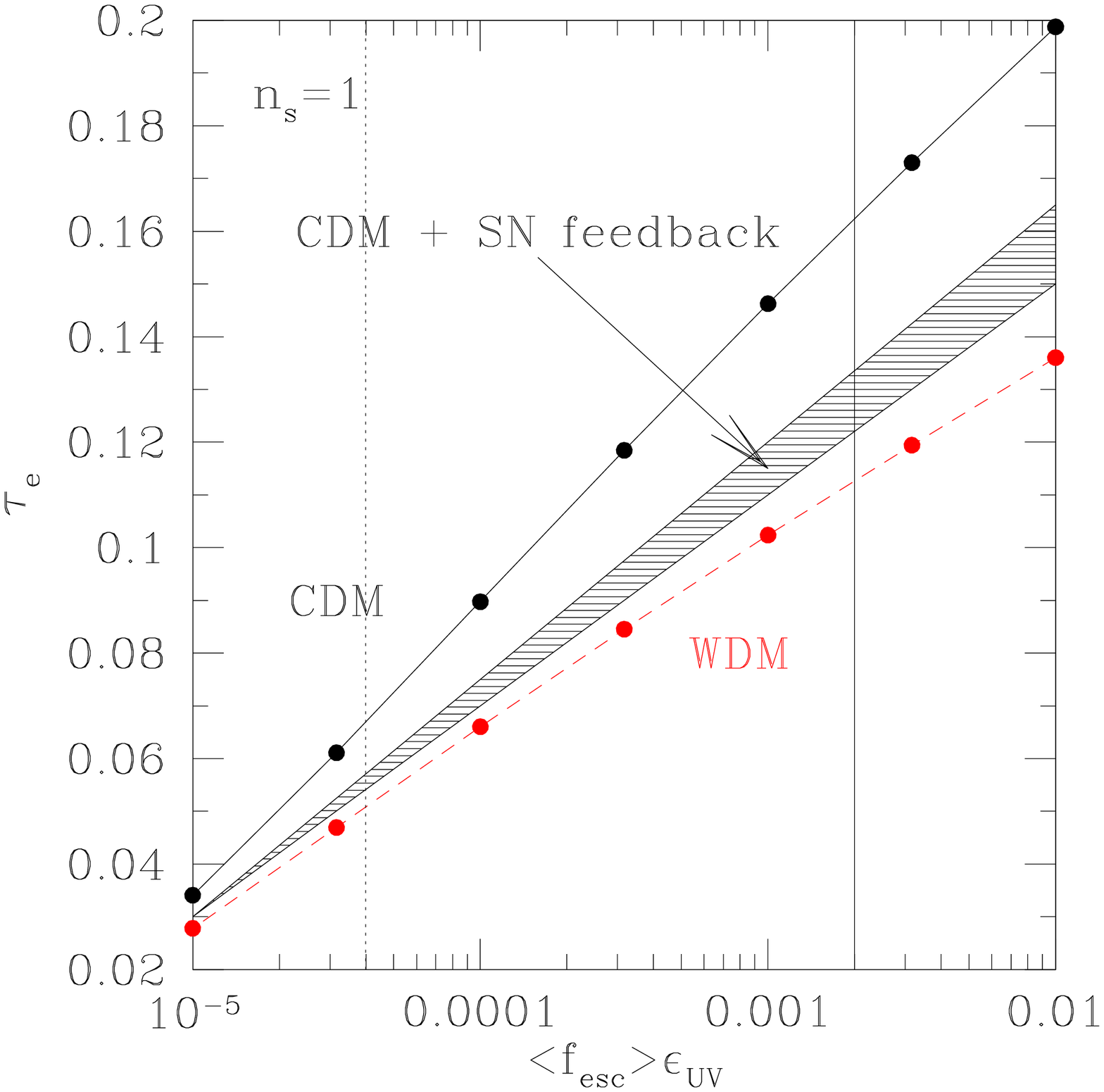,height=8cm}}
\caption{\label{fig:wdm} Optical depth to Thomson scattering, \taue,
  as a function of the effective efficiency of UV emission \fesc\euv
  for the CDM (solid) and a WDM (dashed) with particle mass 1.25 MeV
  cosmologies. We use a semianalytic model with free parameters chosen
  to reproduce the results of the numerical simulation and the
  observed SFR at $z \sim 6$. Here feedback effects are neglected. We
  estimate that they would reduce \taue by 20-25\%, as shown in the
  shaded region for the CDM model, and leave the WDM model unaltered.
  The vertical solid line shows the maximum effective efficiency
  available from thermonuclear sources. Plausible (\popII) models
  would shift it to the dotted line reducing \taue to the level of
  7-8\% which is the value directly inferred by observations of quasar
  reionisation \protect{\citep{Fan:03, ChiuFO:03}} and is below the
  $2\sigma$ error bar of WMAP.}
\end{figure}

In this section we use the semianalytic code to study the dependence
of \taue on the cosmology \citep[\eg,][]{Yoshida:03a, Yoshida:03b}. In
particular we focus on models with a power spectrum of perturbations
that differs in power at small-scales.  Here we study the dependence
of \taue on the slope of the initial power spectrum of perturbations
for a CDM and warm dark matter (WDM) cosmology. The results presented
in this section are intended to show the dependence of \taue on the
cosmological parameters, but they are not sophisticated enough to give
a precise estimate of \taue. The reader should keep in mind that in
these models the most important feedback processes - those due to SNe
- are neglected. In \fig~\ref{fig:pk} we show \taue as a function of
the slope, $n_s$, of the power spectrum of initial density
perturbations.  We normalise the power spectrum according to the
amplitude $A$ at wave number $k=0.05$ h Mpc$^{-1}$ measured by WMAP
\citep{Verde:03}.  Consequently the variance $\sigma_8$ at $z=0$ is
determined by the value of $n_s$.  We show the results for values of
the product \fesc$\epsilon_{\rm UV}=\epsilon_{\rm UV}^{\rm
  eff}=10^{-2}, 10^{-3}$ and $10^{-4}$.  The shaded contours show the
$1\sigma$ confidence limits in the \taue-$n_s$ plane derived by the
WMAP analysis of temperature fluctuations of the CMB
\citep{Spergel:03, Kogut:03}. A small tilt, $n_s<1$, suppresses the
power on small scales delaying star formation in the first galaxies
and the redshift of reionisation, as also found by \cite{ChiuFO:03}.
We find a linear relation between \taue and the spectral index $n_s$:
\[
\tau_e \simeq \tau_e(n_s=1)4.54(n_s-0.78),
\]
where $\tau_e(n_s=1)$ is the value of \taue for $n_s=1$. In
\fig~\ref{fig:wdm} we show $\tau_e(n_s=1)$ as a function of
$\epsilon_{\rm UV}^{\rm eff}$ for the $\Lambda$CDM and $\Lambda$WDM
cosmology. The optical depth \taue increases exponentially with
$\epsilon_{\rm UV}^{\rm eff}$:
\[
\tau_e(n_s=1)=a+b\log{\left({\epsilon_{\rm UV}^{\rm eff} \over 10^{-3}}\right)},
\]
where $a=0.145(0.10)$ and $b=0.055(0.036)$ for the $\Lambda$CDM
($\Lambda$WDM) cosmology. Our results roughly agree with a similar
study by \cite{ChiuFO:03}, but a precise comparison with their work is
not possible since we adopt a constant effective efficiency of UV
emission, $\epsilon_{\rm UV}^{\rm eff}$, while they adopt a time
dependent effective efficiency.

The value of \taue$=0.13$ found in our numerical simulation with weak
SN feedback (128L1VMb) is shown as a circle in \fig~\ref{fig:pk}. The
corresponding value found in the semianalytic model, shown with a
square, is about $15$\% larger. This can be due to the finite
resolution of the numerical simulation or the radiative and weak SN
explosion feedbacks included in the numerical simulation but,
considering these difference in the underlying assumptions, the
agreement between the two methods is encouraging. In the numerical
simulations including strong feedback, the SN explosion heating
reduces the value of \taue by $20-25$\% as compared to the weak
feedback case.

We have also examined a warm dark matter model with our semianalytic
set of calculations.  If the strong SN feedback seen in the numerical
simulations produces a similar reduction of \taue in the semianalytic
CDM model, we find that the warm dark matter model with particle mass
$1.25$ keV does, as expected, produce a lower \taue but the effect is
surprisingly small and estimated to be on $10$ \% for a given value of
$\epsilon_{\rm UV}^{\rm eff}$ (\cf, \fig~\ref{fig:wdm}). The effect of
SN explosions will leave the WDM model essentially unchanged because
small-mass galaxies are not present in this model.  The
photoevaporation of minihaloes is expected to reduce \taue in CDM
models, but this effect is not present in WDM models.

\section{Conclusion}\label{sec:con}

We have used cosmological hydro-dynamical simulations with radiative
and SN explosion feedback to study the reionisation of the IGM by
\pop3 stars. Simple calculations have been used to argue that a large
\taue is difficult to achieve and is incompatible with low metal
enrichment of the IGM/ISM if \pop3 stars end their lives as
pair-instability SNe or smaller-mass hypernovae.  A semianalytic code
is used to study the dependence of \taue on cosmological parameters.

The main results from the cosmological simulations are summarised in
the following points.
\begin{enumerate}
\item It is very difficult or impossible to use stellar sources to
  reionise the universe at \taue$>0.13$ given a maximum efficiency of
  ionising radiation production typical (extreme \pop3) of
  thermonuclear fusion reactions $\epsilon^{max}_{\rm UV}=2 \times
  10^{-3}$ and \fesc$=0.5$.  When feedback from SN explosions is
  included, the limited mass resolution of the cosmological
  simulations is not crucial for achieving convergence for the global
  star formation history in the first galaxies. However a larger box
  size is required because the galaxies that contribute mostly to star
  formation are those that are the most massive.  Our studies of
  convergence indicate that increasing the mass resolution over what
  we utilised in the 128L2VM run (which gave \taue$=0.13$) would not
  increase \taue for models with SN feedback.
\item If metal production is normal (top-heavy IMF, or pair
  instability SNe or hypernovae), the mechanical energy input by SN
  explosions is about ten times larger than for Salpeter IMF. This
  produces strong outflows in galaxies with masses $M_{\rm dm} \simlt
  10^{9}$ M$_\odot$, reducing their star formation and delaying
  reionisation to \zrei$< 10$.
\item If the ratio of metal to ionising photons from metal poor \pop3
  stars is normal, then metal enrichment of the ISM prevents
  metal-poor stars from being produced for long enough to reionise the
  IGM. One requires a factor of $10^3$ reduction of yield or
  efficiency of mixing to alter this conclusion.
\item The issues raised in the previous two points can be alleviated
  if most \pop3 stars collapse into a BH without producing a large
  amount of metals. This solves the metal pollution problem making the
  epoch of \pop3 star domination longer. And in this scenario the
  mechanical feedback from SN/hypernovae explosions would also be
  reduced.
\item If metal poor stars are initially important, then the secondary
  production of ionising radiation due to BH accretion may be
  more important than the primary production.
\item As a byproduct of the semianalytic treatment we find that a warm
  dark matter model with particle mass $1.25$ keV does, as expected,
  produce a lower \taue, but the effect is surprisingly small, and
  estimated to be on 10 \% for a given value of $\epsilon_{\rm
    UV}^{\rm eff}$.
\item If we (self consistently) neglect \pop3 and have more normal
  \popII star formation properties we derive a much lower \taue of
  $0.07-0.08$ consistent with other authors but inconsistent with WMAP
  results.
\end{enumerate}

In conclusion, if the first stars were massive and most of the
radiation escaped from the host galaxies, the increase of \taue could
be within the $1\sigma$ confidence limits of WMAP data. But these two
conditions are not sufficient. The energy liberated by SN explosions
and the metal yields would have to have been much smaller than
expected.  This requirement seems to agree nicely with one possible
interpretation of the abundance pattern in the recently discovered
most iron-deficient star, HE0107-5240, and other metal-poor but
carbon-rich stars.  Subluminous supernova explosions with $E_{51} \sim
0.3$ are characterised by a large fall-back on the central black hole
and their ejecta are carbon rich and iron deficient. The gas, enriched
in carbon and oxygen, can cool fast enough to produce a second
generation of stars with abundance patterns similar to HE0107-5240
\citep{UmedaN:03}.  In addition the abundance pattern of HE0107-5240
is not compatible with the yields from pair-instability SNe.  An
interesting suggestion is that the energy of a SN might be low if the
BH remnant is not spinning, and large for spinning BHs.  Note that
these conclusions are true even if the first stars were not
supermassive.  Indeed, the abundance pattern of HE0107-5240 can be
explained assuming a mass of 25 M$_\odot$ for the SN progenitor which
is roughly the mass indicated by \cite{Abel:02} calculation of ``the
first star''. An interesting consequence of this scenario is a copious
production of rather massive BHs from the first stars.  For this
reason, the secondary radiation from accretion on seed BHs might be a
more important source of ionising radiation than the primary radiation
from the massive stars \citep[see also,][]{MadauR:03}.  BH accretion
quickly builds up an X-ray background that can keep the IGM partially
ionised before the complete reionisation by stellar sources at $z \sim
7$. We investigate this scenario in separate papers (paper~IIa and
paper~IIb).

\subsection*{ACKNOWLEDGEMENTS}
M.R. is supported by a PPARC theory grant. Research conducted in
cooperation with Silicon Graphics/Cray Research utilising the Origin
3800 supercomputer (COSMOS) at DAMTP, Cambridge.  COSMOS is a UK-CCC
facility which is supported by HEFCE and PPARC. M.R. thanks Martin
Haehnelt and the European Community Research and Training Network
``The Physics of the Intergalactic Medium'' for support. The authors
would like to thank Renyue Cen, Andrea Ferrara, Nick Gnedin and Martin
Rees for useful conversations and the anonymous referee for useful
suggestions that improved the manuscript. M.R would like to thank
Erika Yoshino for proofreading the manuscript and support.

\bibliographystyle{/home/ricotti/Latex/TeX/apj}
\bibliography{/home/ricotti/Latex/TeX/archive}

\begin{thebibliography}{}

\bibitem[\protect\citeauthoryear{{Abel}, {Bryan}, \& {Norman}}{{Abel}
  et~al.}{2002}]{Abel:02}
{Abel}, T., {Bryan}, G.~L.,  \& {Norman}, M.~L. 2002, Science, 295, 93

\bibitem[\protect\citeauthoryear{{Barkana} \& {Loeb}}{{Barkana} \&
  {Loeb}}{2002}]{Barkana:02}
{Barkana}, R.,  \& {Loeb}, A. 2002, \apj, 578, 1

\bibitem[\protect\citeauthoryear{{Bean}, {Melchiorri}, \& {Silk}}{{Bean}
  et~al.}{2003}]{BeanS:03}
{Bean}, R., {Melchiorri}, A.,  \& {Silk}, J. 2003, \prd, 68, 083501

\bibitem[\protect\citeauthoryear{{Becker} et~al.}{{Becker}
  et~al.}{2001}]{Becker:01}
{Becker}, R.~H., et~al. 2001, \aj, 122, 2850

\bibitem[\protect\citeauthoryear{{Bennett} et~al.}{{Bennett}
  et~al.}{2003}]{Bennet:03}
{Bennett}, C.~L., et~al. 2003, \apjs, 148, 1

\bibitem[\protect\citeauthoryear{{Benson} et~al.}{{Benson}
  et~al.}{2001}]{Benson:01}
{Benson}, A.~J., {Nusser}, A., {Sugiyama}, N.,  \& {Lacey}, C.~G. 2001, \mnras,
  320, 153

\bibitem[\protect\citeauthoryear{{Bond}, {Carr}, \& {Hogan}}{{Bond}
  et~al.}{1986}]{Bond:86}
{Bond}, J.~R., {Carr}, B.~J.,  \& {Hogan}, C.~J. 1986, \apj, 306, 428

\bibitem[\protect\citeauthoryear{{Bromm}, {Coppi}, \& {Larson}}{{Bromm}
  et~al.}{1999}]{BrommCL:99}
{Bromm}, V., {Coppi}, P.~S.,  \& {Larson}, R.~B. 1999, \apjl, 527, L5

\bibitem[\protect\citeauthoryear{{Bromm} et~al.}{{Bromm}
  et~al.}{2001}]{BrommF:01}
{Bromm}, V., {Ferrara}, A., {Coppi}, P.~S.,  \& {Larson}, R.~B. 2001, \mnras,
  328, 969

\bibitem[\protect\citeauthoryear{{Cen}}{{Cen}}{2003a}]{Cen:03b}
{Cen}, R. 2003a, \apjl, 591, L5

\bibitem[\protect\citeauthoryear{{Cen}}{{Cen}}{2003b}]{Cen:03a}
{Cen}, R. 2003b, \apj, 591, 12

\bibitem[\protect\citeauthoryear{{Cen} \& {Ostriker}}{{Cen} \&
  {Ostriker}}{1992}]{CenO:92}
{Cen}, R.,  \& {Ostriker}, J.~P. 1992, \apjl, 399, L113

\bibitem[\protect\citeauthoryear{{Cen} \& {Ostriker}}{{Cen} \&
  {Ostriker}}{1999}]{CenO:99}
{Cen}, R.,  \& {Ostriker}, J.~P. 1999, \apjl, 519, L109

\bibitem[\protect\citeauthoryear{{Chiu}, {Fan}, \& {Ostriker}}{{Chiu}
  et~al.}{2003}]{ChiuFO:03}
{Chiu}, W.~A., {Fan}, X.,  \& {Ostriker}, J.~P. 2003, \apj, 599, 759

\bibitem[\protect\citeauthoryear{{Chiu} \& {Ostriker}}{{Chiu} \&
  {Ostriker}}{2000}]{Chiu:00}
{Chiu}, W.~A.,  \& {Ostriker}, J.~P. 2000, \apj, 534, 507

\bibitem[\protect\citeauthoryear{{Christlieb} et~al.}{{Christlieb}
  et~al.}{2002}]{Christlieb:02}
{Christlieb}, N., et~al. 2002, Nature, 419, 904

\bibitem[\protect\citeauthoryear{{Ciardi}, {Ferrara}, \& {White}}{{Ciardi}
  et~al.}{2003}]{CiardiFW:03}
{Ciardi}, B., {Ferrara}, A.,  \& {White}, S.~D.~M. 2003, \mnras, 344, L7

\bibitem[\protect\citeauthoryear{{Ciardi}, {Stoehr}, \& {White}}{{Ciardi}
  et~al.}{2003}]{CiardiSW:03}
{Ciardi}, B., {Stoehr}, F.,  \& {White}, S.~D.~M. 2003, \mnras, 343, 1101

\bibitem[\protect\citeauthoryear{{Couchman} \& {Rees}}{{Couchman} \&
  {Rees}}{1986}]{CouchmanR:86}
{Couchman}, H.~M.~P.,  \& {Rees}, M.~J. 1986, \mnras, 221, 53

\bibitem[\protect\citeauthoryear{{Fan} et~al.}{{Fan} et~al.}{2003}]{Fan:03}
{Fan}, X., et~al. 2003, \aj, 125, 1649

\bibitem[\protect\citeauthoryear{{Fukugita}, {Hogan}, \& {Peebles}}{{Fukugita}
  et~al.}{1998}]{Fukugita:98}
{Fukugita}, M., {Hogan}, C.~J.,  \& {Peebles}, P.~J.~E. 1998, \apj, 503, 518

\bibitem[\protect\citeauthoryear{{Gebhardt} et~al.}{{Gebhardt}
  et~al.}{2000}]{Gebhardt:00}
{Gebhardt}, K., et~al. 2000, \apjl, 539, L13

\bibitem[\protect\citeauthoryear{{Gnedin}}{{Gnedin}}{1995}]{Gnedin:95}
{Gnedin}, N.~Y. 1995, \apjs, 97, 231

\bibitem[\protect\citeauthoryear{{Gnedin}}{{Gnedin}}{1998}]{Gnedin:98a}
{Gnedin}, N.~Y. 1998, \mnras, 294, 407

\bibitem[\protect\citeauthoryear{{Gnedin}}{{Gnedin}}{2000}]{Gnedin:00}
{Gnedin}, N.~Y. 2000, \apj, 535, 530

\bibitem[\protect\citeauthoryear{{Gnedin} \& {Abel}}{{Gnedin} \&
  {Abel}}{2001}]{GnedinA:01}
{Gnedin}, N.~Y.,  \& {Abel}, T. 2001, New Astronomy, 6, 437

\bibitem[\protect\citeauthoryear{{Gnedin}, {Ostriker}, \& {Rees}}{{Gnedin}
  et~al.}{1995}]{GnedinOR:95}
{Gnedin}, N.~Y., {Ostriker}, J.~P.,  \& {Rees}, M.~J. 1995, \apj, 438, 40

\bibitem[\protect\citeauthoryear{{Guedens}, {Clancy}, \& {Liddle}}{{Guedens}
  et~al.}{2002}]{Guedens:02}
{Guedens}, R., {Clancy}, D.,  \& {Liddle}, A.~R. 2002, \prd, 66, 43513

\bibitem[\protect\citeauthoryear{{Haiman}, {Abel}, \& {Madau}}{{Haiman}
  et~al.}{2001}]{Haiman:01}
{Haiman}, Z., {Abel}, T.,  \& {Madau}, P. 2001, \apj, 551, 599

\bibitem[\protect\citeauthoryear{{Hansen} \& {Haiman}}{{Hansen} \&
  {Haiman}}{2004}]{HansenH:03}
{Hansen}, S.~H.,  \& {Haiman}, Z. 2004, \apj, 600, 26

\bibitem[\protect\citeauthoryear{{Islam}, {Taylor}, \& {Silk}}{{Islam}
  et~al.}{2003a}]{Islam:03a}
{Islam}, R.~R., {Taylor}, J.~E.,  \& {Silk}, J. 2003a, ArXiv Astrophysics
  e-prints

\bibitem[\protect\citeauthoryear{{Islam}, {Taylor}, \& {Silk}}{{Islam}
  et~al.}{2003b}]{Islam:03b}
{Islam}, R.~R., {Taylor}, J.~E.,  \& {Silk}, J. 2003b, ArXiv Astrophysics
  e-prints

\bibitem[\protect\citeauthoryear{{Kogut} et~al.}{{Kogut}
  et~al.}{2003}]{Kogut:03}
{Kogut}, A., et~al. 2003, \apjs, 148, 161

\bibitem[\protect\citeauthoryear{{Kormendy} \& {Richstone}}{{Kormendy} \&
  {Richstone}}{1995}]{Kormendy:95}
{Kormendy}, J.,  \& {Richstone}, D. 1995, \araa, 33, 581

\bibitem[\protect\citeauthoryear{{Lanzetta} et~al.}{{Lanzetta}
  et~al.}{2002}]{Lanzetta:02}
{Lanzetta}, K.~M., {Yahata}, N., {Pascarelle}, S., {Chen}, H.,  \&
  {Fern{\'a}ndez-Soto}, A. 2002, \apj, 570, 492

\bibitem[\protect\citeauthoryear{{Larson}}{{Larson}}{1998}]{Larson:98}
{Larson}, R.~B. 1998, \mnras, 301, 569

\bibitem[\protect\citeauthoryear{{Leitherer} et~al.}{{Leitherer}
  et~al.}{1999}]{Leitherer:99}
{Leitherer}, C., et~al. 1999, \apjs, 123, 3

\bibitem[\protect\citeauthoryear{{Madau} \& {Rees}}{{Madau} \&
  {Rees}}{2001}]{MadauR:01}
{Madau}, P.,  \& {Rees}, M.~J. 2001, \apjl, 551, L27

\bibitem[\protect\citeauthoryear{{Madau} et~al.}{{Madau}
  et~al.}{2003}]{MadauR:03}
{Madau}, P., {Rees}, M.~J., {Volonteri}, M., {Haardt}, F.,  \& {Oh}, S.~P.
  2003, submitted (astro-ph/0310223)

\bibitem[\protect\citeauthoryear{{Madau} \& {Shull}}{{Madau} \&
  {Shull}}{1996}]{Madau:96}
{Madau}, P.,  \& {Shull}, J.~M. 1996, \apj, 457, 551

\bibitem[\protect\citeauthoryear{{Mateo}}{{Mateo}}{1998}]{Mateo:98}
{Mateo}, M.~L. 1998, \araa, 36, 435

\bibitem[\protect\citeauthoryear{{Miralda-Escud{\' e}}}{{Miralda-Escud{\'
  e}}}{2003}]{Miralda:03}
{Miralda-Escud{\' e}}, J. 2003, Science, 300, 1904

\bibitem[\protect\citeauthoryear{{Nakamura} \& {Umemura}}{{Nakamura} \&
  {Umemura}}{1999}]{Nakamura:99}
{Nakamura}, F.,  \& {Umemura}, M. 1999, \apj, 515, 239

\bibitem[\protect\citeauthoryear{{Nakamura} \& {Umemura}}{{Nakamura} \&
  {Umemura}}{2001}]{Nakamura:01}
{Nakamura}, F.,  \& {Umemura}, M. 2001, \apj, 548, 19

\bibitem[\protect\citeauthoryear{{Oh}, {Cooray}, \& {Kamionkowski}}{{Oh}
  et~al.}{2003}]{Oh:03}
{Oh}, S.~P., {Cooray}, A.,  \& {Kamionkowski}, M. 2003, \mnras, 342, L20

\bibitem[\protect\citeauthoryear{{Omukai}}{{Omukai}}{2000}]{Omukai:00}
{Omukai}, K. 2000, \apj, 534, 809

\bibitem[\protect\citeauthoryear{{Omukai} \& {Palla}}{{Omukai} \&
  {Palla}}{2003}]{OmukaiP:03}
{Omukai}, K.,  \& {Palla}, F. 2003, \apj, 589, 677

\bibitem[\protect\citeauthoryear{{Ostriker} \& {Gnedin}}{{Ostriker} \&
  {Gnedin}}{1996}]{OstrikerG:96}
{Ostriker}, J.~P.,  \& {Gnedin}, N.~Y. 1996, \apjl, 472, L63

\bibitem[\protect\citeauthoryear{{Persic} \& {Salucci}}{{Persic} \&
  {Salucci}}{1992}]{Persic:92}
{Persic}, M.,  \& {Salucci}, P. 1992, \mnras, 258, 14P

\bibitem[\protect\citeauthoryear{{Ricotti}}{{Ricotti}}{2003a}]{Ricotti:03}
{Ricotti}, M. 2003a, \mnras, 344, 1237

\bibitem[\protect\citeauthoryear{{Ricotti}}{{Ricotti}}{2003b}]{Ricotti_pr:03}
{Ricotti}, M. 2003b, in The IGM/Galaxy Connection: The Distribution of Baryons
  at z=0, ASSL Conference Proceedings Vol. 281. Edited by Jessica L. Rosenberg
  and Mary E. Putman. Kluwer Academic Publishers, Dordrecht, 2003, p.193, 193

\bibitem[\protect\citeauthoryear{{Ricotti}, {Gnedin}, \& {Shull}}{{Ricotti}
  et~al.}{2001}]{RicottiGS:01}
{Ricotti}, M., {Gnedin}, N.~Y.,  \& {Shull}, J.~M. 2001, \apj, 560, 580

\bibitem[\protect\citeauthoryear{{Ricotti}, {Gnedin}, \& {Shull}}{{Ricotti}
  et~al.}{2002a}]{RicottiGSa:02}
{Ricotti}, M., {Gnedin}, N.~Y.,  \& {Shull}, J.~M. 2002a, \apj, 575, 33

\bibitem[\protect\citeauthoryear{{Ricotti}, {Gnedin}, \& {Shull}}{{Ricotti}
  et~al.}{2002b}]{RicottiGSb:02}
{Ricotti}, M., {Gnedin}, N.~Y.,  \& {Shull}, J.~M. 2002b, \apj, 575, 49

\bibitem[\protect\citeauthoryear{{Ricotti} \& {Ostriker}}{{Ricotti} \&
  {Ostriker}}{2003}]{RicottiO:03}
{Ricotti}, M.,  \& {Ostriker}, J.~P. 2003, submitted, (astro-ph/0310331),
  (paper~IIa)

\bibitem[\protect\citeauthoryear{{Ricotti}, {Ostriker}, \& {Gnedin}}{{Ricotti}
  et~al.}{2004}]{RicottiOG:03}
{Ricotti}, M., {Ostriker}, J.~P.,  \& {Gnedin}, N.~Y. 2004, in preparation,
  (paper~IIb)

\bibitem[\protect\citeauthoryear{{Salvaterra} \& {Ferrara}}{{Salvaterra} \&
  {Ferrara}}{2003}]{Salvaterra:03}
{Salvaterra}, R.,  \& {Ferrara}, A. 2003, \mnras, 339, 973

\bibitem[\protect\citeauthoryear{{Santos}, {Bromm}, \& {Kamionkowski}}{{Santos}
  et~al.}{2002}]{Santos:02}
{Santos}, M.~R., {Bromm}, V.,  \& {Kamionkowski}, M. 2002, \mnras, 336, 1082

\bibitem[\protect\citeauthoryear{{Schaye} et~al.}{{Schaye}
  et~al.}{2003}]{Schaye:03}
{Schaye}, J., {Aguirre}, A., {Kim}, T., {Theuns}, T., {Rauch}, M.,  \&
  {Sargent}, W.~L.~W. 2003, \apj, 596, 768

\bibitem[\protect\citeauthoryear{{Schneider} et~al.}{{Schneider}
  et~al.}{2002}]{Schneider:02}
{Schneider}, R., {Ferrara}, A., {Natarajan}, P.,  \& {Omukai}, K. 2002, \apj,
  571, 30

\bibitem[\protect\citeauthoryear{{Sciama}}{{Sciama}}{1982}]{Sciama:82}
{Sciama}, D.~W. 1982, \mnras, 198, 1P

\bibitem[\protect\citeauthoryear{{Shapiro} \& {Raga}}{{Shapiro} \&
  {Raga}}{2000}]{Shapiro:00}
{Shapiro}, P.~R.,  \& {Raga}, A.~C. 2000, in Revista Mexicana de Astronomia y
  Astrofisica Conference Series, 292

\bibitem[\protect\citeauthoryear{{Shchekinov}}{{Shchekinov}}{1986}]{Shchekinov%
:86}
{Shchekinov}, I.~A. 1986, Astrofizika, 24, 579

\bibitem[\protect\citeauthoryear{{Sokasian} et~al.}{{Sokasian}
  et~al.}{2003}]{Sokasian:03}
{Sokasian}, A., {Yoshida}, N., {Abel}, T., {Hernquist}, L.,  \& {Springel}, V.
  2003, ArXiv Astrophysics e-prints

\bibitem[\protect\citeauthoryear{{Somerville} \& {Livio}}{{Somerville} \&
  {Livio}}{2003}]{Somerville:03}
{Somerville}, R.~S.,  \& {Livio}, M. 2003, \apj, 593, 611

\bibitem[\protect\citeauthoryear{{Spergel} et~al.}{{Spergel}
  et~al.}{2003}]{Spergel:03}
{Spergel}, D.~N., et~al. 2003, \apjs, 148, 175

\bibitem[\protect\citeauthoryear{{Uehara} et~al.}{{Uehara}
  et~al.}{1996}]{Uehara:96}
{Uehara}, H., {Susa}, H., {Nishi}, R., {Yamada}, M.,  \& {Nakamura}, T. 1996,
  \apjl, 473, L95

\bibitem[\protect\citeauthoryear{{Umeda} \& {Nomoto}}{{Umeda} \&
  {Nomoto}}{2003}]{UmedaN:03}
{Umeda}, H.,  \& {Nomoto}, K. 2003, Nature, 422, 871

\bibitem[\protect\citeauthoryear{{Vandenberg}, {Stetson}, \&
  {Bolte}}{{Vandenberg} et~al.}{1996}]{Vandenberg:96}
{Vandenberg}, D.~A., {Stetson}, P.~B.,  \& {Bolte}, M. 1996, \araa, 34, 461

\bibitem[\protect\citeauthoryear{{Venkatesan} \& {Truran}}{{Venkatesan} \&
  {Truran}}{2003}]{VenkatesanT:03}
{Venkatesan}, A.,  \& {Truran}, J.~W. 2003, \apjl, 594, L1

\bibitem[\protect\citeauthoryear{{Verde} et~al.}{{Verde}
  et~al.}{2003}]{Verde:03}
{Verde}, L., et~al. 2003, \apjs, 148, 195

\bibitem[\protect\citeauthoryear{{Wada} \& {Venkatesan}}{{Wada} \&
  {Venkatesan}}{2003}]{WadaV:03}
{Wada}, K.,  \& {Venkatesan}, A. 2003, \apj, 591, 38

\bibitem[\protect\citeauthoryear{{Wyithe} \& {Loeb}}{{Wyithe} \&
  {Loeb}}{2003}]{WyitheL:03}
{Wyithe}, J.~S.~B.,  \& {Loeb}, A. 2003, \apjl, 588, L69

\bibitem[\protect\citeauthoryear{{Yoshida} et~al.}{{Yoshida}
  et~al.}{2003a}]{Yoshida:03b}
{Yoshida}, N., {Sokasian}, A., {Hernquist}, L.,  \& {Springel}, V. 2003a, \apj,
  598, 73

\bibitem[\protect\citeauthoryear{{Yoshida} et~al.}{{Yoshida}
  et~al.}{2003b}]{Yoshida:03a}
{Yoshida}, N., {Sokasian}, A., {Hernquist}, L.,  \& {Springel}, V. 2003b,
  \apjl, 591, L1

\end{thebibliography}

\label{lastpage}
\end{document}